# Spin-polarized Superconductivity and High-Chern Insulators in Twisted Rhombohedral Graphene Family

Zihao Huo[1]*, Zexu Li[1]*, Wenxuan Wang[1]*†, Gengdong Zhou[1], Qiu Yang[1], Xin Sui[1,2], Zaizhe Zhang[1], Kenji Watanabe[3], Takashi Taniguchi[4], Zhida Song[1], Kaihui Liu[2,5†], and Xiaobo Lu[1,5†]

[1]International Center for Quantum Materials, School of Physics, Peking University, Beijing 100871, China
[2]State Key Laboratory for Mesoscopic Physics, Frontiers Science Centre for Nano-optoelectronics, School of Physics, Peking University, Beijing 100871, China
[3]Research Center for Electronic and Optical Materials, National Institute for Material Science, 1-1 Namiki, Tsukuba 305-0044, Japan.
[4]Research Center for Materials Nanoarchitectonics, National Institute for Material Science, 1-1 Namiki, Tsukuba 305-0044, Japan
[5]Interdisciplinary Institute of Light-Element Quantum Materials and Research Centre for Light-Element Advanced Materials, Peking University, Beijing 100407, China.

*These authors contributed equally to this work.
†E-mail:wenxuanwang@pku.edu.cn; khliu@pku.edu.cn; xiaobolu@pku.edu.cn;

**Rhombohedral multilayer graphene (RMG) has emerged as a remarkably versatile platform for exploring strong-correlation-driven quantum states arising from low-energy topological flat bands. When reconstructed by the moiré superlattice, these bands host a wide range of emergent novel states, including integer and fractional Chern insulators and unconventional superconductivity (SC). Here, we firstly report the simultaneous emergence of widespread spin-polarized SC and high Chern insulators (HCIs) in twisted bilayer-multilayer RMG system (2+n, n=4,5,6). The SC states in (2+n) system exhibit different responses to the in-plane magnetic field $B_{||}$, with SC being suppressed, enhanced and induced by $B_{||}$. The latter two are consistent with spin-triplet pairing. Along with SC, angle- and layer-dependent HCIs with tunable Chern numbers emerge. Moreover, the fractional HCI in 2+4 system survives under high B|| which can induce SC in the same device. Our work not only establishs twisted bilayer–multilayer rhombohedral graphene as a unified platform for studying SC and HCIs, but also opens a pathway towards multiple co-propagating chiral Majorana channels by coupling spin-polarized SC to HCIs.**

The interplay between topological order and superconductivity(SC) stands as a central challenge in condensed matter physics, particularly in the quest for topological SC. Over the past decade, inducing SC into topological states relied on the proximity effect in heterostructures, interfacing conventional *s*-wave superconductors either with materials hosting strong spin-orbit coupling (SOC) and Zeeman spin splitting[1–5], or with chiral edge states in quantum anomalous Hall effect (QAHE) systems[6–10]. However, a fundamental paradox hinders this approach: the chiral edge states in QAHE are typically spin-polarized, whereas most superconductors provide spin-singlet Cooper pairs. The spin mismatch between conventional spin-singlet Cooper pairs and spin-polarized chiral edge states can hinder efficient proximity coupling, which may be facilitated by spin-mixing mechanisms such as strong SOC[10–12]. A theoretically elegant, yet experimentally elusive, resolution to this bottleneck is the employment of spin-triplet superconductors[13–17]. With equal-spin pairing, a triplet superconductor can naturally match the spin polarization of the QAHE edge states, enabling spin-conserving superconducting proximity coupling without relying on additional spin-mixing mechanisms. Furthermore, unlike conventional (C = 1) Chern insulators with a single chiral edge channel, high-Chern-number phases support multiple chiral edge channels that can participate in superconducting proximity coupling[18–20]. Coupling these channels to spin-triplet superconductivity may provide a route towards topological superconducting states with multiple co-propagating chiral Majorana modes[7,21–26].

Due to the profound theoretical significance, an intrinsic material platform capable of concurrently hosting and electrostatically tuning both robust triplet SC and large-Chern-number topological phases has been deeply coveted. Rhombohedral multilayer graphene (RMG) has emerged as a versatile platform for unconventional superconductivity[27–43], including states with signatures of spin-polarized or chiral pairing. In parallel, RMG/hBN superlattice and twisted RMG host a rich family of integer and fractional Chern insulators[41–54], including high-Chern-number insulators (HCIs)[55–59]. Despite these advances, the coexistence of spin-polarized SC and HCIs within the same electrically tunable system remains largely unexplored.

Here, we report the simultaneous emergence of spin-polarized SC and large Chern number QAHE in twisted RMG systems composed of Bernal bilayer graphene (BBG) and RMG, denoted by 2+n where n is the layer number of rhombohedral component. The HCIs exhibit clear tunability with twist angle and layer number; the SC states, emerging against a complex isospin-polarization background, display a rich evolution under the in-plane magnetic field and show compelling evidence for spin-triplet pairing.

## Tunable high Chern insulators

Fig. 1a illustrates the device architecture of twisted bilayer–multilayer RMG. A small twist angle between the BBG and RMG creates a moiré superlattice, enabling the engineering of the electronic bands. We find that the Chern number is not only modulated by the RMG layer number n but also by the moiré twist angle $\theta$. Fig. 1b presents the calculated single-particle band structures of the (2+4) system for two representative twist angles, 1.2° and 1.4°. Under an interlayer potential difference $\Delta_U = -14$ meV (corresponding to displacement field $D > 0$), the first conduction band becomes flatter compared to the zero-field case ($\Delta_U = 0$) and exhibits non-trivial topology. At 1.2°, the flat band carries a Chern number C = 3, which changes to C = 4 as the twist angle is raised to 1.4°. Note that the $\Delta_U$ in the non-interacting Hamiltonian is treated as an approximation to the displacement field experimentally controlled by dual gates. A quantitative mapping between $\Delta_U$ and $D$ requires further self-consistent electrostatic calculations. Experimentally, we confirm this tunability of the topological Chern number through both twist angle $\theta$ and layer number n. In Device 2+4_1 ($\theta = 1.38°$), displacement-field-dependent HCIs with C = 3 and C = 4 emerge at $\nu = 1$, with the zero-field quantized C = 4 HCI being the dominant state[55]. By contrast, Device 2+4_2 ($\theta = 1.18°$) exhibits a predominant C = 3 HCI in the corresponding parameter regime. Fig. 1c shows the symmetrized longitudinal resistivity $\rho_{xx}$ and antisymmetrized Hall resistivity $\rho_{xy}$ phase diagrams as functions of $\nu$ and $D$ near $\nu = 1$, measured at perpendicular magnetic field $B_\perp = \pm 0.2$ T. Furthermore, we performed magnetic transport measurements along specific $D$ field, yielding Landau fan diagrams (Fig. 1d). According to the Středa relation, $\partial n/\partial B_\perp = C \cdot e/h$, the expected C = 3 trajectory at $\nu = 1$, traced by the evolution of the $\rho_{xx}$ minimum and the $h/3e^2$-quantized plateau in $\rho_{xy}$, extrapolates continuously to zero magnetic field, providing a key signature of the C = 3 Chern insulator. Magnetic hysteresis measurements (Fig. 1e) further demonstrate the QAHE, with the C = 3 state reaching 99% quantization.

Nevertheless, in Device 2+5 ($\theta$ = 1.40°), the HCI at $\nu$ = 1 carries a Chern number C = 4. The pronounced zero-resistance minimum in $\rho_{xx}$ and the $h/4e^2$-quantized Hall plateau in $\rho_{xy}$ establish the realization of a zero-field C = 4 HCI (Fig. 1f, Extended Data Fig. 3a). Landau fan diagrams (Fig. 1g, Extended Data Fig. 3b-c) and magnetic hysteresis loops (Fig. 2e) further verify the C = 4 HCI state. Linecuts along $\nu$ = 1 (Extended Data Fig. 3d) demonstrate that this state remains robust over a broad displacement field range of $D/\varepsilon_0 \approx$ 0.55~0.68 V/nm. This observation differs from the previously established design principle of C = n in the twisted monolayer graphene/RMG family[58,59], indicating that the topological band engineering enabled by integrating BBG with RMG is governed by a more complex mechanism beyond the layer-number dependence alone. Further theoretical investigations are required to fully understand this additional degree of freedom.

**Pervasive spin-polarized superconductivity**

More strikingly, systematic measurements reveal the coexistence of integer/fractional HCIs and SC phases within the same device. We adopt the following terminology throughout this work: all states exhibiting distinct SC critical behaviour are referred to as SC states, with the prefix "SC-like (SCL)" named for those that saturate at a finite resistance and "SC" for those that reach zero resistance. Possible mechanisms underlying the finite resistance are considered in the Methods. In Device 2+4_1 ($\theta$ = 1.38°), we investigate the evolution of the phase diagram at different in-plane magnetic field $B_{||}$ = 0 T, 1 T, and 2 T, and summarize the resulting phase revolution in a schematic $\nu$-$D$ phase diagram (Fig. 2a). At $B_{||}$ = 0 T, Device 2+4_1 only exhibits cascades of HCIs around $\nu$ = 2/3, 1 and 3 (Fig. 2c). Upon applying a finite $B_{||}$, distinct SC regions emerge near the phase boundaries in different regions of the phase diagram. Features of integer and fractional HCIs remains robust under different $B_{||}$, demonstrated in Fig. 2d-e and Extended Data Fig. 1. Fig. 2b shows a representative line-cut that traverses both SCL1 and QAH regions. Taking SCL1 at $B_{||}$ = 1 T as an example (other SC regions shown in Extended Data Fig. 2 and Supplementary Fig. 1), SCL1 appears in the vicinity of the high-resistance insulating state induced by finite $D$ at the charge neutrality point (CNP) of the RMG system (Fig. 2f). A pronounced SC region emerges and progressively expands in $\nu$ range with increasing $B_{||}$, eventually reaching a nearly field-independent extent up to experimental limit of 2 T (Fig. 2g and Supplementary Fig. 1). Temperature-dependent resistance measurements, together with the characteristic spindle-shaped patterns in the differential resistance d$V$/d$I$ maps, provide strong evidence for the superconductivity nature of SCL1. The Berezinskii–Kosterlitz–Thouless (BKT) transition temperature $T_{BKT}$ of SCL1 is estimated to be about 110 mK from temperature-dependent measurements of nonlinear d$V$/d$I$ versus DC bias current $I_{DC}$, where the characteristic relation $V_{xx} \propto I_{DC}^3$ is observed (Fig. 2j). Fig. 2k compares the temperature-dependent resistance curves for SCL1–3 at $B_{||}$ = 1 T. At base temperature, SCL3 and SCL4 do not reach a fully developed SC phase, but the spindle-shaped patterns still support their incipient superconductivity nature. Moreover, Device 2+4_2 ($\theta$ = 1.18°) also exhibits the coexistence of a C = 3 HCI state and two SC states, shown in Extended Data Fig. 8. SCL1 emerges at $B_{||}$ = 0 T and enhanced by increasing $B_{||}$, while SCL2 is induced above $B_{||} \approx$ 0.5 T. Such behavior is qualitatively similar to that observed in Device 2+5, as discussed below.

Measurements of Device 2+5 ($\theta$ = 1.40°) under different in-plane fields have been summarized in Fig. 3a. Not only a C = 4 HCI but also numerous SC regions were observed at $B_{||}$ = 0 T (Fig.3c). With increasing in-plane field, certain SC regions are suppressed, others are expanded and enhanced, with additional phases emerging at higher fields. Remarkably, the major SC regions are located along the phase boundaries separating different isospin-polarized states, a characteristic feature widely observed in RMG systems. The dominant SC regions can be classified into two branches, denoted SCL River A and SCL River B, according to their distinct evolution in the phase diagram. SCL River A extends approximately parallel to $V_{tg}$, consistent with strong top-gate screening, whereas SCL River B shows an overall $V_{tg}$-parallel evolution, with its trajectory progressively bending as it encounters competing moiré insulating states. Such unconventional response to the dual gate may imply their possible surface-state-related origin[35,36,42]. Additionally, SCL C and D emerge only under higher in-plane fields. At $B_{\parallel} = 0$ T (Fig. 3d), SCL River A can be further subdivided into three patches, SCL A1-A3. SCL River B, in contrast, contains only SCL B1 and B5, along with not-fully-developed SCL B2 and B7 (Supplementary Fig. 4). Fig. 3e shows temperature-dependent resistance measurements along the yellow dashed line in Fig. 3d, revealing the progressive suppression of SC with increasing temperature. Representative $R$-$T$ curves extracted at selected points (Fig. 3f) further characterize the SC states. The SC states are further distinguished by characteristic spindle-shaped features in the d$V$/d$I$ maps as a function of $I_{DC}$ and $B_{\perp}$ (Fig. 3g and Supplementary Fig. 3). Combining these measurements with the temperature-dependent nonlinear d$V$/d$I$ characteristics allows the BKT transition temperatures to be extracted for selected SC states. For instance, SCL A2 gives $T_{BKT} \approx 100$ mK.

Applying an in-plane magnetic field strongly reshapes the SC landscape of Device 2+5. At $B_{\parallel} = 1$ T, both SCL Rivers A and B expand substantially across the phase diagram (Fig. 4a,b). While SCL A3 and B7 are suppressed, SCL A1 and A2 become connected through the newly emerging SC A4. Within SCL River B, SCL B1, B2 and B5 are markedly enhanced by $B_{\parallel}$, whereas SC B3, SCL B4 and B6 emerge near the original phase boundaries upon introducing $B_{\parallel}$. The d$V$/d$I$ measurement along the dashed trajectory in Fig. 4a further resolves the distinct current responses of SCL A1, A2 and SC A4 (Fig. 4c). The field evolution is particularly evident near SCL A3: the $R_{xx}$ map as a function of $V_{bg}$ and $B_{\parallel}$ shows that SCL A3 is suppressed at $B_{\parallel} \approx 0.2$ T, whereas SC A4 emerges at $B_{\parallel} \approx 0.8$ T (Fig. 4d). With further increasing $B_{\parallel}$, SC A4 at this position is gradually replaced by SCL A1, an evolution further resolved by d$V$/d$I$ measurements as functions of $I_{DC}$ and $B_{\parallel}$ at the center of the SC A4 region (Fig. 4e). Furthermore, SCL River A evolves similarly upon increasing temperature and reducing $B_{\parallel}$ (Fig. 4f and Supplementary Fig. 5), with SC progressively suppressed in both cases.

A similarly rich field evolution occurs in SCL River B. As $B_{\parallel}$ is increased, SCL B2 and SC B3 emerge near the original phase boundary (Fig. 4g,h). Tracking the center of SC B3 reveals its gradual replacement by SCL B2 with increasing $B_{\parallel}$ (Fig. 4i). The temperature-dependent measurements further establish the SC character of SCL B2 and SC B3, with the extracted $R$–$T$ curves yielding critical temperatures ranging from 70 to 130 mK (Fig. 4j,k). For the newly emerged SC A4 at $B_{\parallel} = 1$ T, BKT analysis yields $T_{BKT} \approx 75$ mK (Fig. 4l). Finally, the characteristic spindle-shaped features in the d$V$/d$I$ maps as functions of $I_{DC}$ and $B_{\perp}$ provide further transport evidence for the SC nature of SC A4 and B3 (Fig. 4m,n). In comparison, the 2+6 system also exhibits a similar phenomenology (Extended Data Fig. 9). Pronounced correlated insulating states emerge at $\nu = -8/3$ and $-4/3$, whearas an anomalous Hall effect develops at $\nu = 2/3$. This behavior may reflect the formation of charge-density-wave order. Nonetheless, a long SCL river emerges upon applying $B_{\parallel}$. Taken together, the pervasive $B_{\parallel}$-induced or enhanced SC observed across these devices highlights its robustness to variations in device quality and layer number.

Finally, we characterized the response of these SC states to a perpendicular magnetic field. Taking the SC B3 at $B_{\parallel}$ = 1 T in Device 2+5 as a representative example, we first measured $R_{xx}$ as a function of $B_{\perp}$ over a range of temperatures (Fig. 5a). At each temperature, the critical perpendicular field $B_c$ was defined as the field at which $R_{xx}$ reaches 70% of the normal-state resistance, with the uncertainty estimated using thresholds of 65% and 75%. We then fit the temperature dependence of $B_c$ (Fig. 5b) to the Ginzburg–Landau (GL) theory, $B_c = (\Phi_0 / (2\pi\xi^2)) (1 - T/T_c)$, to extract the zero-temperature GL coherence length $\xi_{GL(T=0K)}$. For SC B3 at $B_{\parallel}$ = 1 T, $\xi_{GL(T=0K)}$ is approximately 160 nm, substantially exceeding the average inter-carrier distance in this system. Since the SC regions are more extensive and pronounced at $B_{\parallel}$ = 1 T, we systematically extracted $\xi_{GL(T=0K)}$ at representative positions for each SC state under this field, and summarize these results in a scatter plot (Fig. 5c). Comparing with the average inter-carrier distance, coherence lengths for most regions are substantially larger, in some cases by more than two orders of magnitude. This indicates that the SC states observed in our 2+n system tends to be governed by a weak-coupling pairing mechanism, similar to that observed in thick RMG systems[35,42].

## Discussion

Two primary pieces of evidence support the interpretation that a substantial part of the SC states observed in the twisted bilayer-multilayer RMG system favors spin-triplet pairing. First, their response to in-plane magnetic fields is difficult to reconcile with conventional Pauli-limited spin-singlet SC. An in-plane magnetic field introduces Zeeman splitting between opposite-spin electrons and therefore tends to suppress spin-singlet pairing. In contrast, most of the SC states observed here are either enhanced or induced by $B_{\parallel}$, and persist to fields well beyond the estimated Pauli limit (Methods), behavior that is naturally compatible with spin-triplet pairing. Similar behaviors of SC states have been observed previously in BBG[60] and RMG system[28,31,33–36,42]. We attribute these $B_{\parallel}$-induced or enhanced SC to the van Hove singularities in RMG and speculate such SC behavior to be ubiquitous in sufficiently clean RMG systems. Besides, the moiré effect mainly affects SC by suppressing competing phases (e.g., the LAF state and the displacement-field-induced insulator near CNP), thus creating band conditions more conducive to SC.

Second, the evolution of the phase boundaries under $B_{\parallel}$ provides an independent indication that SC is closely associated with a spin-polarized electronic background. In the 2+n systems, the interacting band evolves strongly with tuning parameters such as filling factor $v$ and $D$, making it difficult to obtain well-resolved quantum oscillations for a detailed fermiology analysis. Nevertheless, the in-plane field evolution of the phase diagram remains informative. Taking 2+5 system as an example (Fig. 4d and 4h), SCL Rivers A and B are located on the opposite sides of a polygonal region, outside which is the low-resistance metallic phase. With increasing $B_{\parallel}$, this polygonal region expands outward at the expense of the surrounding metallic phase, behavior consistent with its identification as a spin-polarized phase favored by the in-plane field. Concomitantly, both SCL Rivers A and B extend along the evolving boundaries of this region. The correlated evolution of the putative spin-polarized phase and the adjacent SC states therefore suggests that SC develops in close connection with a spin-polarized electronic background. Together with the anomalous response of SC to $B_{\parallel}$, this phase-boundary evolution provides a second line of evidence favoring spin-triplet pairing. A microscopic understanding of this relationship will require further theoretical investigation.

Beyond providing insight into the microscopic origin of SC in the twisted 2+n systems, our results open opportunities for coupling SC to chiral Chern edge states. Spin-triplet SC can host time-reversal-symmetry-breaking states[26], sharing a compatible symmetry character with zero-field Chern insulators[20]. Together with their spin-polarized nature, this may provide a favorable setting for SC coupling to chiral Chern edge states[13,15–17]. Moreover, the HCIs in the 2+n system support multiple chiral edge channels. The coexistence and electrostatic tunability of these phases motivate the realization of SC–HCI hybrid structures, in which proximity coupling between spin-polarized SC and chiral Chern edge states may provide a route toward exploring multiple co-propagating chiral Majorana modes.

*Note added in proof:* During the preparation of this paper, we became aware of a related study on the anomolous metal in RMG/$WSe_2$ system[40].

## Acknowledgements

X.L. discloses support for the research of this work from the National Natural Science Foundation of China (Grant No. 12521006), the National Key R&D Program (Grant No. 2022YFA1403502 and 2024YFA1409002) and Beijing National Laboratory for Condensed Matter Physics (Grant No. 2025BNLCMPKF001). Z.S. discloses support for the research of this work from National Natural Science Foundation of China (Grant No. 12274005), National Key R&D Program (Grant No. 2021YFA1401900) and Quantum Science and Technology-National Science and Technology Major Project (No. 2021ZD0302403). K.L. discloses support for the research of this work from the National Natural Science Foundation of China (Grant No. 12427806 and 92577201). W.W. discloses support for the research of this work from Quantum Science and Technology-National Science and Technology Major Project (Grant No. 2025ZD0300500) and the Peking University Boya Postdoctoral Fellowship. We thank the Peking Nanofab for process support.

## Author Contributions

X.L., K.L. and W.W. conceived and designed the experiments; Z.H., Z.L.and W.W fabricated the devices and performed the transport measurement with help from others. Z.H., G. Z., Z.L.and W.W., Z.S. and X.L. analyzed the data; G.Z. and Z.S. performed the theoretical modeling; T.T. and K.W. contributed hBN substrates; Z.H. and X.L. wrote the paper with input from others.

## Competing interests

The authors declare no competing interests.

## Data Availability

All data supporting the findings of this study are available within the main text, figures and Supplementary Information, or from the corresponding authors upon request. Source data are provided with this paper.

## Code Availability

Codes that support the findings of this study are available upon request. Codes include scripts for data processing and theoretical modelling.

# Methods

## Device fabrication

The twisted bilayer-multilayer rhombohedral graphene devices were assembled using the standard polycarbonate-based dry transfer procedure with dual hBN encapsulation and graphite gate electrodes, following protocols established in our previous work[55,56]. Prior to stacking, clean and wrinkle-free graphene samples with natural bilayer and multilayer regions were preselected to ensure a consistent initial crystallographic orientation. Layer numbers were identified by optical contrast, and rhombohedral stacking in the tetralayer/pentalayer/hexalayer regions was verified by rapid infrared imaging technique[61] and Raman spectroscopy[62–64]. Selected flakes were then cut into rectangular pieces by laser trimming and subsequently transferred layer by layer in the following sequence: top gate, top hBN, bilayer flake, and multilayer flake, onto a prefabricated substrate. The substrate, consisting of bottom hBN and graphite gate, underwent sequential surface treatment: annealing in an $H_2$(4%)/Ar atmosphere at 350 °C, followed by mechanical cleaning via AFM contact-mode scanning[65] to remove residual contaminants. Post-stacking Raman spectroscopy confirmed the preservation of rhombohedral stacking. Finally, Hall bar geometries were defined by standard electron-beam lithography and reactive-ion etching, with edge contacts formed by evaporated Cr/Au (5/60 nm) electrodes.

## Electrical transport measurement

Standard low-frequency lock-in techniques were employed for all measurements, with devices cooled to a base temperature of about 20 mK in a dilution refrigerator. The dual gates were powered by Keithley 2400 source-meters, and four-terminal resistance measurements were carried out using SR860 lock-in amplifiers, cascaded with SR560 preamplifiers to improve the signal-to-noise ratio.

The carrier density $n$ and electrical displacement field $D$ were defined as $n = (c_{tg}V_{tg} + c_{bg}V_{bg})/e$ and $D/\varepsilon_0 = (c_{tg}V_{tg} - c_{bg}V_{bg})/(2\varepsilon_0)$, where $V_{tg}$ and $V_{bg}$ correspond to the top and bottom gate voltage, respectively, $c_{tg}$ and $c_{bg}$ are the capacitance of the top and bottom gates per unit area, and $\varepsilon_0$ is the permittivity of vacuum. Moiré filling factor $\nu$ and twist angle $\theta$ were comprehensively calibrated and verified by positions of correlated insulators, quantum Hall effects as well as Brown-Zak oscillation. Alignment angle ($\theta$) between bilayer and tetralayer/pentalayer graphene is calculated by $\lambda = a/\sqrt{2(1-\cos\theta)}$ where lattice constant of graphene $a$ = 0.246nm.

The magnetic field for low-temperature measurements was supplied by a vector magnet, with maximum fields of 6 T for $B_z$ and 2 T for $B_x$. Owing to a slight misalignment angle arising from the mounting of the sample carrier and the fixation of the sample, the applied $B_x$ field is not perfectly parallel to the sample plane. This results in an additional out-of-plane component (typically less than 1%), which is corrected by applying the $B_z$ component during measurement.

## Symmetrization and antisymmetrization

For transport data characterizing the Chern insulator, $\rho_{xx}$ and $\rho_{xy}$ are expected to be symmetrized and antisymmetrized with respect to the magnetic field, respectively, as follows:

$$\rho_{xx}^{\mathrm{Sym}}(B,\nu) = \frac{\rho_{xx}^{\mathrm{Original}}(B,\nu) + \rho_{xx}^{\mathrm{Original}}(-B,\nu)}{2},$$

$$\rho_{xy}^{\mathrm{AntiSym}}(B,\nu) = \frac{\rho_{xy}^{\mathrm{Original}}(B,\nu) - \rho_{xy}^{\mathrm{Original}}(-B,\nu)}{2}.$$

Moreover, for magnetic hysteresis measurement data, the similar symmetrization and antisymmetrization procedures were employed:

$$\rho_{xx,\mathrm{Forward}}^{\mathrm{Sym}}(B) = \frac{\rho_{xx,\mathrm{Forward}}^{\mathrm{Original}}(B) + \rho_{xx,\mathrm{Backward}}^{\mathrm{Original}}(-B)}{2},$$
$$\rho_{xx,\mathrm{Backward}}^{\mathrm{Sym}}(B) = \rho_{xx,\mathrm{Forward}}^{\mathrm{Sym}}(-B),$$
$$\rho_{xy,\mathrm{Forward}}^{\mathrm{AntiSym}}(B) = \frac{\rho_{xy,\mathrm{Forward}}^{\mathrm{Original}}(B) - \rho_{xy,\mathrm{Backward}}^{\mathrm{Original}}(-B)}{2},$$
$$\rho_{xy,\mathrm{Backward}}^{\mathrm{AntiSym}}(B) = -\rho_{xy,\mathrm{Forward}}^{\mathrm{AntiSym}}(-B).$$

**Pauli-limit violation**

According to the weak-coupling BCS theory framework, the Pauli limit $B_P = 1.25\ k_B T_c/\mu_B$ is valid for SC that emerges at zero magnetic field, where $k_B$ is the Boltzmann constant, $\mu_B$ is the Bohr magneton and $T_c$ is the zero-field critical temperature. Evidently, this expression is not directly applicable to SC states induced by finite $B_{\|}$, because $T_c|_{B=0\mathrm{T}}$ is ill-defined in that case. Nevertheless, a conservative estimate can still be performed for a qualitative illustration. Since the observed SC critical temperatures are all below 170 mK, the Pauli limit $B_P = 1.25\ k_B T_c/\mu_B$ gives an upper bound of approximately 316.4 mT for critical in-plane fields, which is far below the critical in-plane fields of these SC states—some regions are not fully developed or do not even appear at this field. For SCL A3 and B7 in Device 2+5, however, the critical in-plane fields are below the Pauli limit, and these states should therefore not be spin-triplet superconductors.

**Discussion on the finite resistance observed in SC-like states.**

Several SC-like (SCL) regions in our devices exhibit a pronounced resistance decrease upon cooling and characteristic nonlinear differential-resistance responses, but remain resistive down to the lowest accessible temperatures. This behavior contrasts with superconducting regions in which the resistance approaches zero, motivating our phenomenological distinction between SCL and fully developed SC states. First, the finite-resistance SCL states exhibit robust and reproducible characteristics upon changing the measurement configuration and the current direction (Extended Data Fig. 10), including a clear distinction from the zero-resistance states. This robustness suggests that the observed finite resistance is unlikely to be primarily associated with sample inhomogeneity. Besides, among SCL states, some show no clear resistance saturation within the experimentally accessible temperature range, whereas others develop a well-defined low-temperature plateau at a finite resistance. (We cannot exclude the possibility that some SCL states could similarly evolve into zero-resistance SC states at temperatures below those accessible in our measurements.) In the following, we focus on the latter SCL states and discuss its possible connection to anomalous metal (AM) transport[40].

Similar low-temperature resistance saturation following a SCL transition has been widely discussed in two-dimensional superconductors in the context of AM behavior, where substantial SC correlations may survive without the establishment of a globally coherent zero-resistance state[66–68]. Within this general framework, the finite resistance may reflect incomplete long-range phase coherence or fluctuations of the SC order parameter. One possible interpretation is a separation between the development of local pairing correlations and global SC coherence. For SC order parameter as $\Delta = |\Delta| e^{i\phi}$, the appearance of a finite local pairing amplitude and the establishment of long-range phase rigidity need not occur simultaneously. Strong fluctuations of $\phi$ can suppress global phase coherence even when substantial local SC correlations remain. If such phase decoherence is accompanied by spatial variations in the local SC susceptibility, the system may develop regions with enhanced local SC correlations whose phases fail to become globally locked.

This spatially inhomogeneous realization of the phase-fluctuation picture is, however, strongly constrained by the characteristic length scales in the present system. In theoretical models of SC puddles embedded in a metallic background, individual SC regions must be sufficiently large to sustain a locally developed order parameter. The corresponding critical puddle radius is of order the SC coherence length[68], $R_c \sim \xi$. This requirement places a strong constraint on a simple moiré-scale puddle interpretation in the present system. The Ginzburg–Landau coherence lengths extracted from representative SC states in our work are typically of order 100−300 nm. By comparison, the primary moiré wavelength is only approximately 11 nm, making a granular Josephson network based on individual moiré-scale SC regions unlikely.

A spatially inhomogeneous SC picture has recently been discussed in twisted trilayer graphene[69–71], where a longer-wavelength supermoiré modulation may induce spatial variations in the local SC susceptibility and thereby separate local SC from global Josephson-mediated coherence. By contrast, neither our (2+x) devices nor the recently studied RMG/$WSe_2$ system[40] exhibits an identified structural modulation on a length scale comparable to the relevant SC coherence length. In this respect, the RMG/WSe2 system provides a closer phenomenological comparison to our Device 2+5. In both Device 2+5 and RMG/WSe2, a finite-resistance SCL state appears in close proximity to a zero-resistance SC state, with the two regimes exhibiting markedly different responses to both perpendicular and in-plane magnetic fields. This phenomenology raises the possibility that the finite- and zero-resistance regimes may involve distinct paired states, with different susceptibilities to magnetic field and other tuning parameters, rather than differing solely in the degree of global phase coherence. More generally, the relevant fluctuations need not originate from an underlying structural modulation, but may instead involve coupling of the SC order parameter to other collective electronic degrees of freedom.

**Noninteracting continuum model of twisted rhombohedral graphene**

We describe twisted rhombohedral graphene (TRnG) by a continuum model of two ABC-stacked graphene multilayers rotated by relative angles θ, where the top stack contains $N_t$ layers and the bottom stack contains $N_b$ layers. The layer index $\ell = 1,..,N_t + N_b$ are counted from top to bottom. With graphene lattice constant $a$ = 2.46 Å, we define $\mathbf{K}_0 = \left(\frac{4\pi}{3a}, 0\right)$, and $\mathbf{K}_t = R_{\theta/2}\mathbf{K}_0, \mathbf{K}_b = R_{-\theta/2}\mathbf{K}_0$ are the momentum of the Dirac points of monolayer graphenes in the top/bottom multilayers. We also define $\mathbf{q}_1 = \mathbf{K}_b - \mathbf{K}_t, \mathbf{q}_j = R_{\frac{2\pi(j-1)}{3}}\mathbf{q}_1, j = 1,2,3,$ and the Moire lattice constant is $a_M = \frac{4\pi}{3|\mathbf{q}_1|} = \frac{a}{2\sin(\theta/2)}$ where $R_\varphi$ is the anti-clockwise rotation along $z$ direction with angle $\varphi$. We focus on $N_t$ = 2, $N_b$ = 4 case and choose θ = 1.2, 1.4°. In the (A,B) sublattice basis and the stack order (top,bottom), the single-particle Hamiltonian in the positive valley is[72]

$$H_+(\mathbf{k}) = \begin{pmatrix} H_t(\mathbf{k}) & \mathcal{T}^\dagger \\ \mathcal{T} & H_b(\mathbf{k}) \end{pmatrix} + U.$$

For stack s = t,b and each retained reciprocal vector Q, the continuum momentum and the isolated ABC-stack block are defined, in the layer order stated above, by

$$[H_s(\mathbf{k})]_{\mathbf{Q}\ell,\mathbf{Q}'\ell'} = \delta_{\mathbf{Q}\mathbf{Q}'}\{\delta_{\ell\ell'}h_D(\mathbf{p_Q}) + \delta_{\ell,\ell'+1}t(\mathbf{p_Q}) + \delta_{\ell',\ell+1}t^\dagger(\mathbf{p_Q}) \\ +\delta_{\ell,\ell'+2}t' + \delta_{\ell',\ell+2}t'^\dagger\}, \mathbf{p_Q} = \mathbf{k} + \mathbf{Q}, s = t, b.$$

Here $h_D$ is the monolayer Dirac Hamiltonian, $t(\mathbf{p})$ couples layer $\ell'$ to layer $\ell'$ + 1, and $t'$ couples layer $\ell'$ to layer $\ell'$ + 2; explicitly,

$$h_D(\mathbf{p}) = v_f \begin{pmatrix} 0 & p_- \\ p_+ & 0 \end{pmatrix}, p_\pm = p_x \pm ip_y, \\ t(\mathbf{p}) = \begin{pmatrix} -v_4p_- & -v_3p_+ \\ t_1 & -v_4p_- \end{pmatrix}, t' = \begin{pmatrix} 0 & t_2 \\ 0 & 0 \end{pmatrix}.$$

The displacement field is represented by sublattice-independent layer potentials

$$[U(\ell)]_{\mathbf{Q}\ell,\mathbf{Q}'\ell'} = \Delta_U\left(\ell - N_t - \frac{1}{2}\right)\delta_{\mathbf{Q}\mathbf{Q}'}\delta_{\ell\ell'}$$

so the two interface layers have potentials $-\Delta_U/2$ and $\Delta_U/2$ and adjacent layers differ by $\Delta_U$. The reciprocal basis are $\mathbf{b}_1 = \mathbf{q}_3 - \mathbf{q}_1, \mathbf{b}_2 = \mathbf{q}_3 - \mathbf{q}_2, \mathbf{G} = m_1\mathbf{b}_1 + m_2\mathbf{b}_2, m_1, m_2 \in \mathbb{Z}, \mathcal{Q}_t = \{\mathbf{G} - \mathbf{q}_1 : |\mathbf{G} - \mathbf{q}_1| < G_c\}, \mathcal{Q}_b = \{\mathbf{G} + \mathbf{q}_1 : |\mathbf{G} + \mathbf{q}_1| < G_c\}$ where G runs over all integer combinations of $\mathbf{b}_1$ and $\mathbf{b}_2$ and $G_c$ is the cutoff. Only the two interface layers are directly coupled between the top and bottom stacks, with[73]

$$T_j = w_0\sigma_0 + w_1[\cos\varphi_j\sigma_x + \sin\varphi_j\sigma_y] = \begin{pmatrix} w_0 & w_1e^{-i\varphi_j} \\ w_1e^{i\varphi_j} & w_0 \end{pmatrix}, \varphi_j = \frac{2\pi(j-1)}{3}, \\ \mathcal{T}_{\mathbf{Q}_b\ell_b,\mathbf{Q}_t\ell_t} = \delta_{\ell_b,1}\delta_{\ell_t,N_t}\sum_{j=1}^{3}\delta_{\mathbf{Q}_b,\mathbf{Q}_t-\mathbf{q}_j}T_j.$$

Unless otherwise specified, the single-particle parameters are chosen as[74] $v_f$ = 6306.2 meV Å, $v_3 = v_4 = 0.04v_f, t_1 = 400\text{meV}, t_2 = 0, w_0 = 88\text{meV}, w_1 = 110\text{meV}$.

# Reference

1. Sau, J. D., Lutchyn, R. M., Tewari, S. & Das Sarma, S. Generic New Platform for Topological Quantum Computation Using Semiconductor Heterostructures. *Phys. Rev. Lett.* **104**, 040502 (2010).
2. Lutchyn, R. M., Sau, J. D. & Das Sarma, S. Majorana Fermions and a Topological Phase Transition in Semiconductor-Superconductor Heterostructures. *Phys. Rev. Lett.* **105**, 077001 (2010).
3. Oreg, Y., Refael, G. & von Oppen, F. Helical Liquids and Majorana Bound States in Quantum Wires. *Phys. Rev. Lett.* **105**, 177002 (2010).
4. Mourik, V. *et al.* Signatures of Majorana Fermions in Hybrid Superconductor-Semiconductor Nanowire Devices. *Science* **336**, 1003–1007 (2012).
5. Das, A. *et al.* Zero-bias peaks and splitting in an Al–InAs nanowire topological superconductor as a signature of Majorana fermions. *Nature Physics* **8**, 887–895 (2012).
6. Fu, L. & Kane, C. L. Superconducting Proximity Effect and Majorana Fermions at the Surface of a Topological Insulator. *Phys. Rev. Lett.* **100**, 096407 (2008).
7. Qi, X.-L., Hughes, T. L. & Zhang, S.-C. Chiral topological superconductor from the quantum Hall state. *Phys. Rev. B* **82**, 184516 (2010).
8. Ii, A., Yada, K., Sato, M. & Tanaka, Y. Theory of edge states in a quantum anomalous Hall insulator/spin-singlet s -wave superconductor hybrid system. *Phys. Rev. B* **83**, 224524 (2011).
9. Wang, J., Zhou, Q., Lian, B. & Zhang, S.-C. Chiral topological superconductor and half-integer conductance plateau from quantum anomalous Hall plateau transition. *Phys. Rev. B* **92**, 064520 (2015).
10. Lian, B., Wang, J. & Zhang, S.-C. Edge-state-induced Andreev oscillation in quantum anomalous Hall insulator-superconductor junctions. *Phys. Rev. B* **93**, 161401 (2016).
11. Kupferschmidt, J. N. & Brouwer, P. W. Andreev reflection at half-metal/superconductor interfaces with nonuniform magnetization. *Phys. Rev. B* **83**, 014512 (2011).
12. Duckheim, M. & Brouwer, P. W. Andreev reflection from noncentrosymmetric superconductors and Majorana bound-state generation in half-metallic ferromagnets. *Phys. Rev. B* **83**, 054513 (2011).
13. Eschrig, M., Kopu, J., Cuevas, J. C. & Schön, G. Theory of Half-Metal/Superconductor Heterostructures. *Phys. Rev. Lett.* **90**, 137003 (2003).
14. Qi, X.-L., Hughes, T. L., Raghu, S. & Zhang, S.-C. Time-Reversal-Invariant Topological Superconductors and Superfluids in Two and Three Dimensions. *Phys. Rev. Lett.* **102**, 187001 (2009).
15. Nakai, R., Nomura, K. & Tanaka, Y. Edge-induced pairing states in a Josephson junction through a spin-polarized quantum anomalous Hall insulator. *Phys. Rev. B* **103**, 184509 (2021).
16. Cheng, Q., Yan, Q. & Sun, Q.-F. Spin-triplet superconductor–quantum anomalous Hall insulator–spin-triplet superconductor Josephson junctions: $0 - \pi$ transition, $\phi$ 0 phase, and switching effects. *Phys. Rev. B* **104**, 134514 (2021).
17. Ohashi, R., Kobayashi, S. & Tanaka, Y. Possible topological phases in quantum anomalous Hall insulator/unconventional superconductor hybrid systems. *Phys. Rev. B* **104**, 134518 (2021).
18. Thouless, D. J., Kohmoto, M., Nightingale, M. P. & Den Nijs, M. Quantized Hall Conductance in a Two-Dimensional Periodic Potential. *Phys. Rev. Lett.* **49**, 405–408 (1982).

19. Avron, J. E., Seiler, R. & Simon, B. Homotopy and Quantization in Condensed Matter Physics. *Phys. Rev. Lett.* **51**, 51–53 (1983).
20. Weng, H., Yu, R., Hu, X., Dai, X. & Fang, Z. Quantum anomalous Hall effect and related topological electronic states. *Advances in Physics* **64**, 227–282 (2015).
21. Li, J. *et al.* Two-dimensional chiral topological superconductivity in Shiba lattices. *Nature Communications* **7**, 12297 (2016).
22. Wang, L. & Wu, M. W. Topological superconductor with a large Chern number and a large bulk excitation gap in single-layer graphene. *Phys. Rev. B* **93**, 054502 (2016).
23. Wang, J. & Lian, B. Multiple Chiral Majorana Fermion Modes and Quantum Transport. *Phys. Rev. Lett.* **121**, 256801 (2018).
24. Lian, B. & Wang, J. Distribution of conductances in chiral topological superconductor junctions. *Phys. Rev. B* **99**, 041404 (2019).
25. He, J. J., Liang, T., Tanaka, Y. & Nagaosa, N. Platform of chiral Majorana edge modes and its quantum transport phenomena. *Communications Physics* **2**, 149 (2019).
26. Ghazaryan, A., Holder, T., Berg, E. & Serbyn, M. Multilayer graphenes as a platform for interaction-driven physics and topological superconductivity. *Phys. Rev. B* **107**, 104502 (2023).
27. Zhou, H., Xie, T., Taniguchi, T., Watanabe, K. & Young, A. F. Superconductivity in rhombohedral trilayer graphene. *Nature* **598**, 434–438 (2021).
28. Seo, J. *et al.* Family of magnetic field-boosted superconductors in rhombohedral graphene. *Nature* 1–3 (2026) doi:10.1038/s41586-026-10815-x.
29. Han, T. *et al.* Signatures of chiral superconductivity in rhombohedral graphene. *Nature* **643**, 654–661 (2025).
30. Kalantre, S. S. *et al.* Fermiology and the Candidate Chiral Superconductor in Rhombohedral Tetralayer Graphene. Preprint at https://doi.org/10.48550/arXiv.2606.05356 (2026).
31. Xie, J. *et al.* Magnetic-Field-Driven Insulator-Superconductor Transition in Rhombohedral Graphene. *Phys. Rev. Lett.* **136**, 176505 (2026).
32. Nguyen, R. Q. *et al.* A Hierarchy of Topological and Superconducting States in Rhombohedral Hexalayer Graphene. Preprint at https://doi.org/10.48550/arXiv.2507.22026 (2025).
33. Deng, J. *et al.* Superconductivity and Ferroelectric Orbital Magnetism in Semimetallic Rhombohedral Hexalayer Graphene. Preprint at https://doi.org/10.48550/arXiv.2508.15909 (2025).
34. Zheng, C. *et al.* Multiple Superconducting Phases in Rhombohedral Heptalayer Graphene. *Chinese Phys. Lett.* **43**, 060718 (2026).
35. Kumar, M. *et al.* Superconductivity from dual-surface carriers in rhombohedral graphene. *Nat. Phys.* https://doi.org/10.1038/s41567-026-03277-5 (2026) doi:10.1038/s41567-026-03277-5.
36. Guo, Y. *et al.* Flat band surface state superconductivity in thick rhombohedral graphene. Preprint at https://doi.org/10.48550/arXiv.2511.17423 (2025).
37. Zhou, Z. *et al.* Competing Orders Driven by Wigner Crystal Phase in Rhombohedral Graphene. Preprint at https://doi.org/10.48550/arXiv.2607.15014 (2026).
38. Yang, J. *et al.* Impact of spin–orbit coupling on superconductivity in rhombohedral graphene. *Nature Materials* **24**, 1058–1065 (2025).
39. Patterson, C. L. *et al.* Superconductivity and spin canting in spin–orbit-coupled trilayer graphene. *Nature* **641**, 632–638 (2025).

40. Okounkova, A. *et al.* Anomalous metal and superconducting phases in rhombohedral graphene. Preprint at https://doi.org/10.48550/arXiv.2607.28425 (2026).
41. Choi, Y. *et al.* Superconductivity and quantized anomalous Hall effect in rhombohedral graphene. *Nature* **639**, 342–347 (2025).
42. Kumar, M. *et al.* Pervasive spin-triplet superconductivity in rhombohedral graphene. Preprint at https://doi.org/10.48550/arXiv.2511.16578 (2025).
43. Zan, X. *et al.* Chern number reversal and emergent superconductivity in rhombohedral graphene induced by in-plane magnetic fields. Preprint at https://doi.org/10.48550/arXiv.2604.27788 (2026).
44. Lu, Z. *et al.* Fractional quantum anomalous Hall effect in multilayer graphene. *Nature* **626**, 759–764 (2024).
45. Lu, Z. *et al.* Extended quantum anomalous Hall states in graphene/hBN moiré superlattices. *Nature* **637**, 1090–1095 (2025).
46. Waters, D. *et al.* Chern Insulators at Integer and Fractional Filling in Moiré Pentalayer Graphene. *Phys. Rev. X* **15**, 011045 (2025).
47. Li, C. *et al.* Stacking-orientation and twist-angle control on integer and fractional Chern insulators in moiré rhombohedral graphene. Preprint at https://doi.org/10.48550/arXiv.2505.01767 (2025).
48. Li, H. *et al.* Competing Chern states revealed by quasiparticle charging in moiré rhombohedral graphene. Preprint at https://doi.org/10.48550/arXiv.2607.08710 (2026).
49. Zheng, C. *et al.* Tunable high-Chern-number Chern insulators in rhombohedral tetralayer graphene/hBN moiré superlattices. Preprint at https://doi.org/10.48550/arXiv.2604.26643 (2026).
50. Xie, J. *et al.* Tunable fractional Chern insulators in rhombohedral graphene superlattices. *Nat. Mater.* **24**, 1042–1048 (2025).
51. Huo, Z. *et al.* Does Moire Matter? Critical Moire Dependence with Quantum Fluctuations in Graphene Based Integer and Fractional Chern Insulators. Preprint at https://doi.org/10.48550/arXiv.2510.15309 (2025).
52. Uzan, M. *et al.* hBN alignment orientation controls moiré strength in rhombohedral graphene. Preprint at https://doi.org/10.48550/arXiv.2507.20647 (2025).
53. Zheng, J. *et al.* Switchable Chern Insulators and Competing Quantum Phases in Rhombohedral Graphene Moiré Superlattices. *Phys. Rev. Lett.* **135**, 136302 (2025).
54. Liu, Q. *et al.* Odd-Chern-Number Quantum Anomalous Hall Effect at Even Filling in Moire Rhombohedral Heptalayer Graphene. *Phys. Rev. Lett.* **136**, 016602 (2026).
55. Li, Z. *et al.* Fractional high-Chern insulator in twisted rhombohedral graphene. *Nature* **655**, 892–898 (2026).
56. Wang, W. *et al.* Programmable Quantum Anomalous Hall Insulator in Twisted Crystalline Flatbands. *Phys. Rev. X* **16**, 011015 (2026).
57. Dong, J. *et al.* Observation of Integer and Fractional Chern insulators in high Chern number flatbands. Preprint at https://doi.org/10.48550/arXiv.2507.09908 (2025).
58. Wang, X. *et al.* High-Chern-number orbital magnetism in twisted rhombohedral graphene. *Nature Materials* 1–7 (2026) doi:10.1038/s41563-026-02659-7.
59. Chen, Z. *et al.* Layer-engineered quantum anomalous Hall effect in twisted rhombohedral graphene. *Nature Materials* 1–7 (2026) doi:10.1038/s41563-026-02711-6.
60. Zhou, H. *et al.* Isospin magnetism and spin-polarized superconductivity in Bernal bilayer graphene. *Science* **375**, 774–778 (2022).

61. Feng, Z. *et al.* Rapid infrared imaging of rhombohedral graphene. *Phys. Rev. Applied* **23**, 034012 (2025).
62. Lui, C. H. *et al.* Imaging Stacking Order in Few-Layer Graphene. *Nano Lett.* **11**, 164–169 (2011).
63. Cong, C. *et al.* Raman Characterization of ABA- and ABC-Stacked Trilayer Graphene. *ACS Nano* **5**, 8760–8768 (2011).
64. Shan, Y. *et al.* Stacking symmetry governed second harmonic generation in graphene trilayers. *Sci. Adv.* **4**, eaat0074 (2018).
65. Lindvall, N., Kalabukhov, A. & Yurgens, A. Cleaning graphene using atomic force microscope. *Journal of Applied Physics* **111**, 064904 (2012).
66. Jaeger, H. M., Haviland, D. B., Orr, B. G. & Goldman, A. M. Onset of superconductivity in ultrathin granular metal films. *Phys. Rev. B* **40**, 182–196 (1989).
67. Eley, S., Gopalakrishnan, S., Goldbart, P. M. & Mason, N. Approaching zero-temperature metallic states in mesoscopic superconductor–normal–superconductor arrays. *Nature Physics* **8**, 59–62 (2012).
68. Kapitulnik, A., Kivelson, S. A. & Spivak, B. *Colloquium* : Anomalous metals: Failed superconductors. *Rev. Mod. Phys.* **91**, 011002 (2019).
69. Zhou, Z. *et al.* Gate-tunable double-dome superconductivity in twisted trilayer graphene. *Nat. Phys.* **21**, 1773–1779 (2025).
70. Xia, L.-Q. *et al.* Magic continuum in multi-moiré twisted trilayer graphene. Preprint at https://doi.org/10.48550/arXiv.2509.03583 (2025).
71. Mahapatra, P. S. *et al.* Quantum criticality and tunable Griffiths phase in superconducting twisted trilayer graphene. Preprint at https://doi.org/10.48550/arXiv.2507.10687 (2026).
72. Herzog-Arbeitman, J. *et al.* Moiré fractional Chern insulators. II. First-principles calculations and continuum models of rhombohedral graphene superlattices. *Phys. Rev. B* **109**, 205122 (2024).
73. Bistritzer, R. & MacDonald, A. H. Moiré bands in twisted double-layer graphene. *Proc. Natl. Acad. Sci. U.S.A.* **108**, 12233–12237 (2011).
74. Zhang, H. *et al.* Moiré enhanced flat band in rhombohedral graphene. *Nature Materials* **25**, 566–572 (2026).

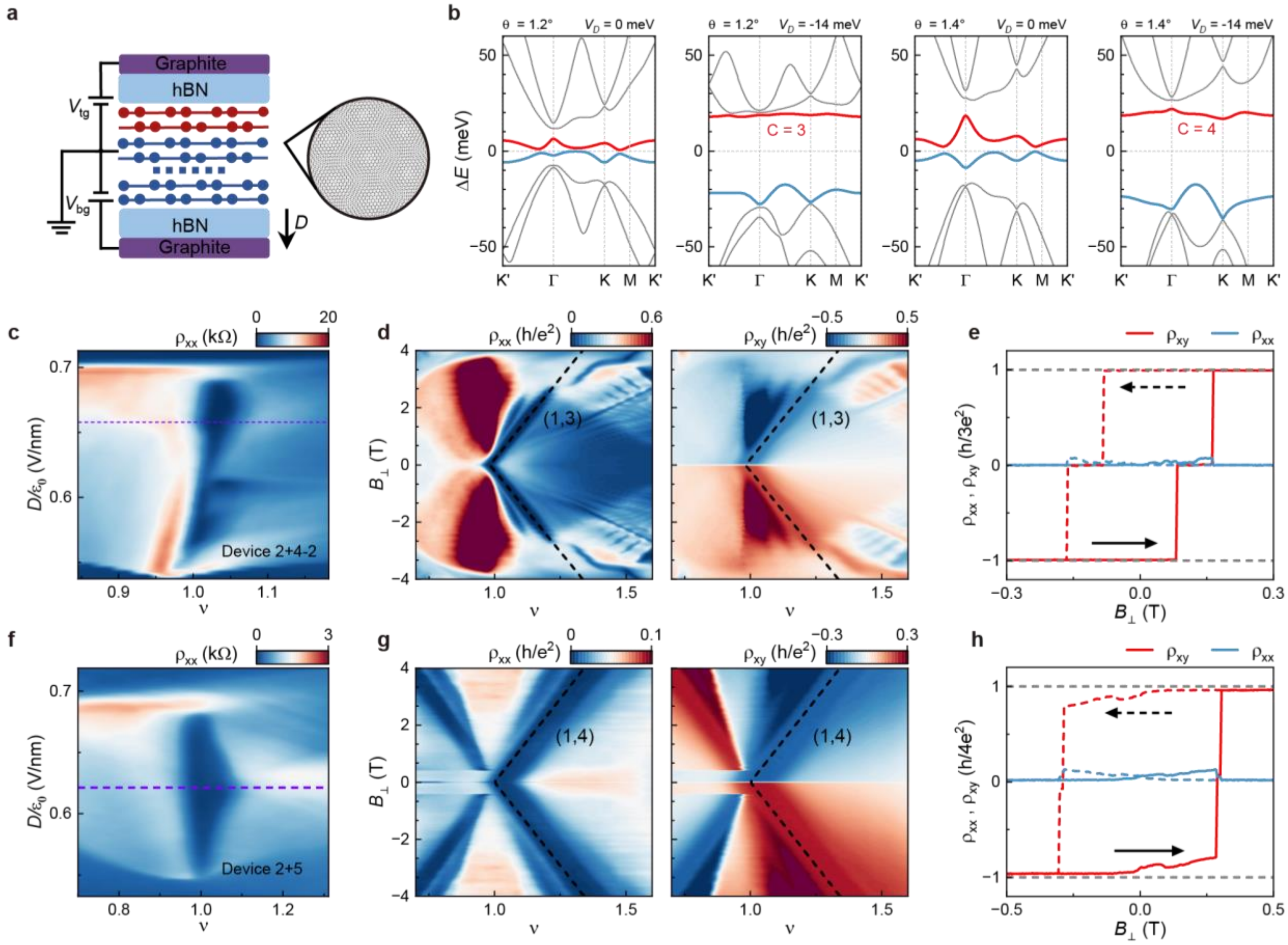

**Fig. 1 | Tunable HCIs in twisted bilayer-multilayer rhombohedral graphene. a,** Schematic of the dual-gate device. Inset, moiré superlattice formed at the interface between the bilayer and multilayer graphene. The displacement field $D$ is defined as positive in the direction from the top gate towards the bottom gate. **b,** Single-particle band structures of twisted 2+4 graphene for twist angle $\theta$ = 1.2°, 1.4° at interlayer potential differences $\Delta_U$ = 0, -14 meV. **c,** Symmetrized longitudinal resistivity $\rho_{xx}$ phase diagram of Device 2+4_2 ($\theta$ = 1.18°) as a function of filling factor $\nu$ and displacement field $D$, measured around $\nu = 1$ at $B_\perp = \pm 0.2$ T. **d,** Landau fan diagrams of symmetrized $\rho_{xx}$ and antisymmetrized $\rho_{xy}$ along $D/\varepsilon_0$ = 0.658 V/nm, indicated by the purple dashed line in **c**. Black dashed lines indicates the C = 3 trajectory emanating from $\nu = 1$ according to the Středa formula, which matches well with the $\rho_{xx}$ minima and the quantized $\rho_{xy}$ plateau. **e,** Magnetic hysteresis loop of symmetrized $\rho_{xx}$ and antisymmetrized $\rho_{xy}$ at $\nu = 1$ for the C=3 HCI, where $\rho_{xy}$ reaches 99% of the quantized value $h/3e^2$. The kinks in the hysteresis loop arise from the symmetrization or antisymmetrization of raw trace with asymmetric coercive fields. **f-h,** Corresponding characterization of the C = 4 HCI in Device 2+5 (θ = 1.40°). **f,** Symmetrized $\rho_{xx}$ as a function of $\nu$ and $D$, obtained at $B_\perp = \pm 0.3$ T. **g,** Landau fan diagrams of symmetrized $\rho_{xx}$ and antisymmetrized $\rho_{xy}$. Black dashed lines mark the C = 4 trajectory from $\nu = 1$. **h,** Magnetic hysteresis of $\rho_{xx}$ and $\rho_{xy}$ for the C = 4 HCI, with the Hall resistance reaching approximately 95% of the quantized value $h/4e^2$.

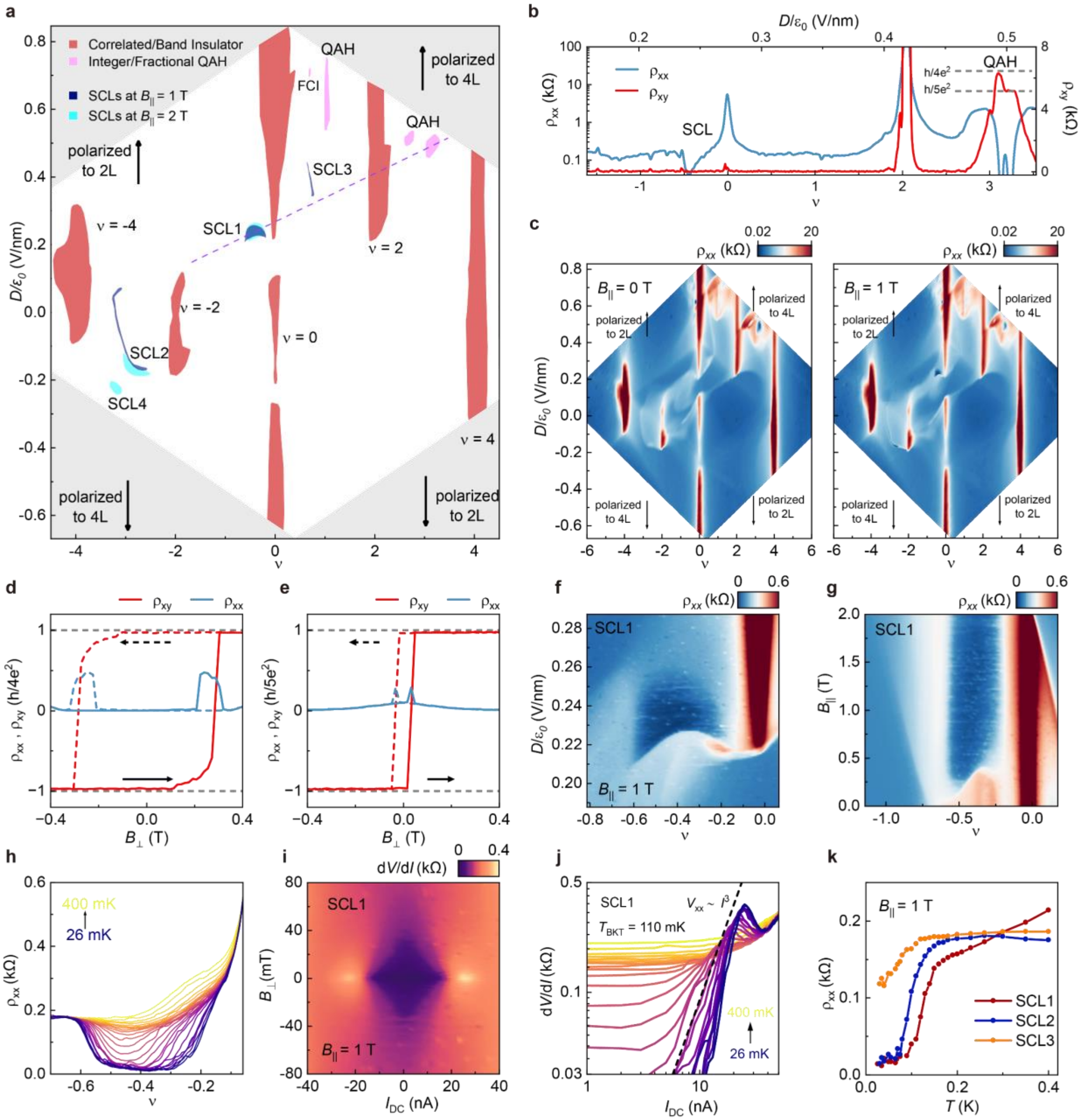

**Fig. 2 | Coexistence of $B_{\parallel}$-induced SC and HCI states in Device 2+4_1. a,** Schematic $\nu$–$D$ phase diagram of Device 2+4_1 ($\theta$ = 1.38°). Blue, pink and red regions denote superconducting, quantum anomalous Hall (QAH) and correlated/band insulating states, respectively. **b,** Longitudinal resistivitiy $\rho_{xx}$ and Hall resistivitiy $\rho_{xy}$ along the purple dashed trajectory in **a,** crossing SCL1 and the C = 4 and C = 5 HCIs. $\rho_{xx}$ (blue) and $\rho_{xy}$ (red) are plotted as functions of $\nu$. **c,** $\rho_{xx}$ as functions of filling factor $\nu$ and displacement field $D$, measured at $B_{\parallel}$ = 0 T (left) and $B_{\parallel}$ = 1 T (right). **d,e,** Magnetic hysteresis of $\rho_{xx}$ and $\rho_{xy}$ as functions of $B_{\perp}$ for the C = 4 (**d**) and C = 5 (**e**) Chern insulators. **f,** Enlarged $\rho_{xx}$ map of SCL1 as a function of $\nu$ and $D$ at $B_{\parallel}$ = 1T. **g,** Evolution of SCL1 with in-plane magnetic field, measured as $\rho_{xx}$ versus of $\nu$ and $B_{\parallel}$ at $D/\varepsilon_0$ = 0.236 V/nm. **h,** Temperature evolution of SCL1, measured as $\rho_{xx}$ versus $\nu$ and $T$ at $B_{\parallel}$ = 1 T. **i,** Differential resistance d$V$/d$I$ map as a function of $I_{DC}$ and $B_{\perp}$ for SCL1. **j,** d$V$/d$I$ as a function of $I_{DC}$ at various temperatures from 26

mK to 400 mK. The red dashed line marks the $V_{xx} \propto I_{DC}^3$ criterion, yielding a Berezinskii–Kosterlitz–Thouless transition temperature $T_{BKT} \approx 110$ mK. **k,** Temperature dependence of $\rho_{xx}$ for SCL1-3 at $B_{\parallel} = 1$ T.

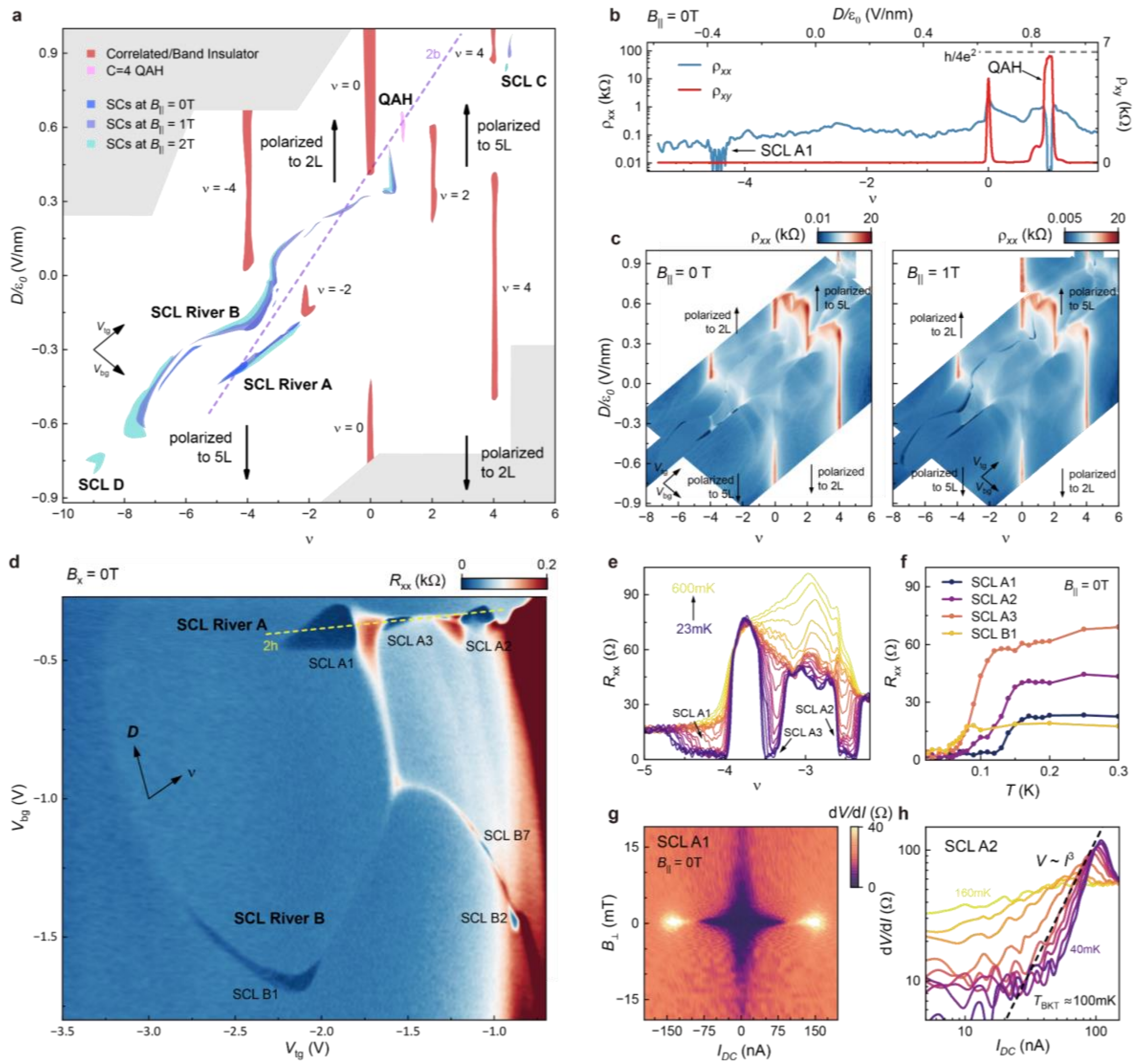

**Fig. 3 | Coexistence of SC and HCI states in Device 2+5. a,** Schematic $\nu$–$D$ phase diagram, summarizing the principal states of Device 2+5 ($\theta = 1.40°$). Colored regions denote distinct states (details labeled in the upper left), and the gray area marks the experimentally inaccessible regime. The arrows indicate the directions of the dual-gate tuning axes. **b,** Longitudinal resistivity $\rho_{xx}$ and Hall resistivity $\rho_{xy}$ along the purple dashed trajectory in **a**, crossing SCL A1 and the C = 4 QAH state near $\nu = 1$. **c,** $\rho_{xx}$ maps of Device 2+5 as functions of $\nu$ and $D$, measured at $B_{\parallel} = 0$ T (left) and $B_{\parallel} = 1$ T (right). **d,** Longitudinal resistance $R_{xx}$ as a function of dual-gate voltage $V_{tg}$ and $V_{bg}$ on the hole doped side. Two main SCL rivers (A and B) can be identified. The arrows indicate the tuning directions of $\nu$ and $D$. **e,** $R_{xx}$ along the yellow dashed trajectory in **d** at a set of temperatures (23–600 mK). **f,** $R$–$T$ characteristics for representative SC regions. The critical temperatures are approximately 75–140 mK. **g,** Differential resistance d$V$/d$I$ of SCL A1 as a function of $I_{DC}$ and $B_{\perp}$, with a critical current of about 140 nA at zero perpendicular field.

**h,** Temperature-dependent d$V$/d$I$ traces of SCL A2 versus $I_{\mathrm{DC}}$. A BKT transition temperature $T_{\mathrm{BKT}}$ of roughly 100 mK is extracted from the fit.

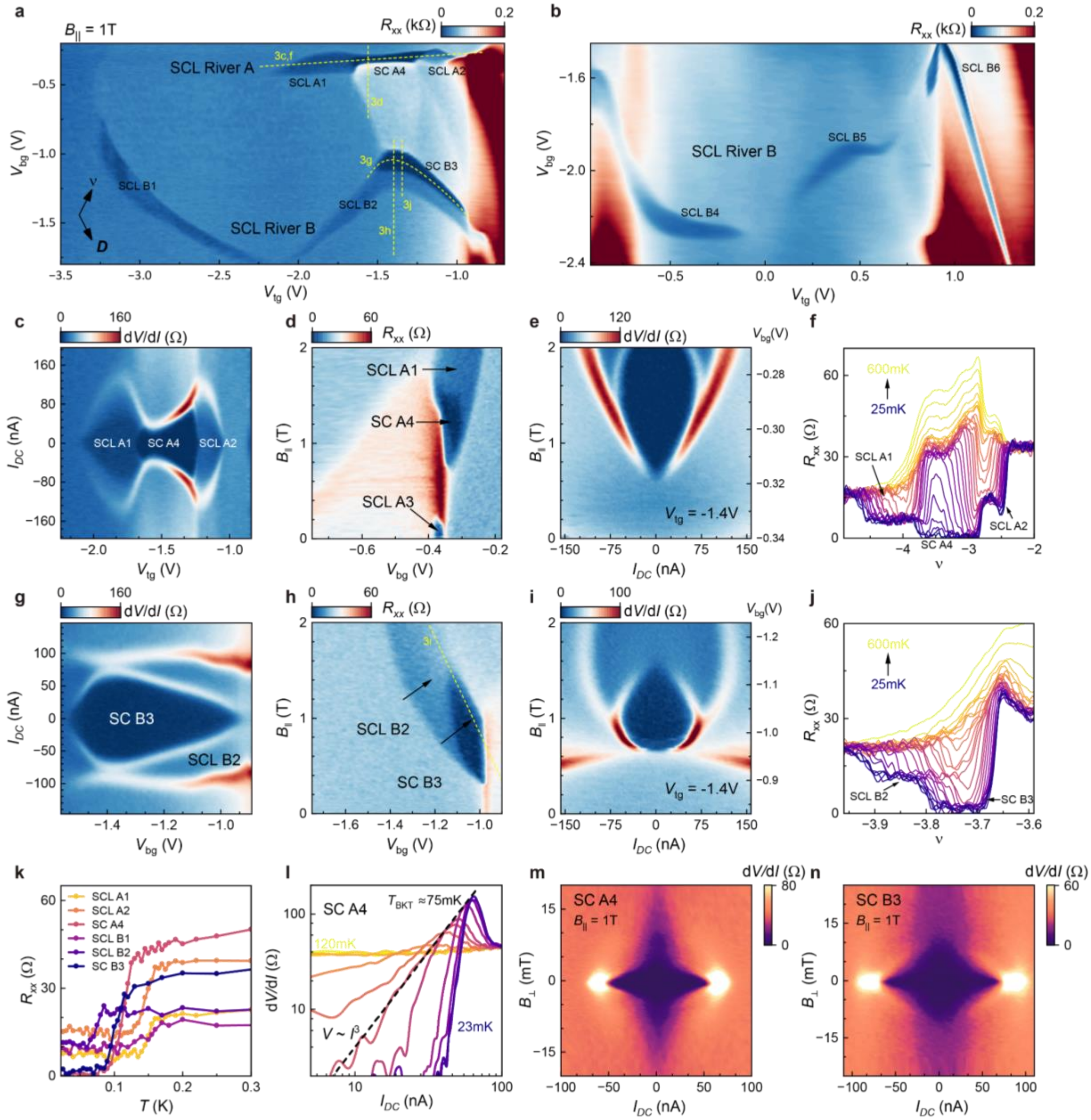

**Fig. 4 | In-plane-field evolution of SC in Device 2+5. a,b,** Longitudinal resistance $R_{\mathrm{xx}}$ of Device 2+5 as a function of dual-gate voltages at $B_{\parallel}$ = 1 T, showing the expanded SC Rivers A and B. **a,** Overview of SCL River A and part of SCL River B. **b,** Enlarged view of SCL River B. Yellow dashed trajectories mark the measurement paths in **c–j**. **c,** Differential resistance d$V$/d$I$ map as a function of top-gate voltage $V_{\mathrm{tg}}$ and DC bias current $I_{\mathrm{DC}}$ along SCL River A at $B_{\parallel}$ = 1 T, resolving SCL A1, SC A4, and SCL A2. **d,** $R_{\mathrm{xx}}$ as a function of bottom-gate voltage $V_{\mathrm{bg}}$ and $B_{\parallel}$ near SCL A3. SCL A3 is suppressed at $B_{\parallel} \approx 0.2$ T, whereas SC A4 emerges at $B_{\parallel} \approx 0.8$ T. **e,** Differential resistance d$V$/d$I$ as a function of $I_{\mathrm{DC}}$ and $B_{\parallel}$ while tracking the centre of SC A4, showing its gradual evolution into SCL A1 at higher field. **f,** Temperature-dependent $R_{\mathrm{xx}}$ line cuts across SCL River A at $B_{\parallel}$ = 1 T, measured from 25 to 600 mK. **g–j,** Analogous measurements to **c–f** performed

on SC B3 and SCL B2. **g,** d$V$/d$I$ map as a function of $V_{bg}$ and $I_{DC}$ across SC B3 and SCL B2 at $B_{\parallel}$ = 1 T. **h,** $R_{xx}$ as a function of $V_{bg}$ and $B_{\parallel}$, showing the field-induced emergence and evolution of SC B3 and SCL B2. **i,** d$V$/d$I$ as a function of $B_{\parallel}$ and $I_{DC}$ while tracking the centre of SC B3. **j,** Temperature-dependent $R_{xx}$ line cuts across SC B3 and SCL B2 at $B_{\parallel}$ = 1 T, measured from 25 to 600 mK. **k,** Temperature dependence of $R_{xx}$ for representative SC states at $B_{\parallel}$ =1 T, with critical temperatures ranging from approximately 70 to 130 mK. **l,** Temperature-dependent d$V$/d$I$ versus temperature at SC A4 for $B_{\parallel}$ = 1T; BKT analysis $T_{BKT}$ ≈ 75mK. **m,n,** d$V$/d$I$ as a function of $I_{DC}$ and $B_{\perp}$ for SC A4 (**m**) and B3 (**n**) at $B_{\parallel}$=1T.

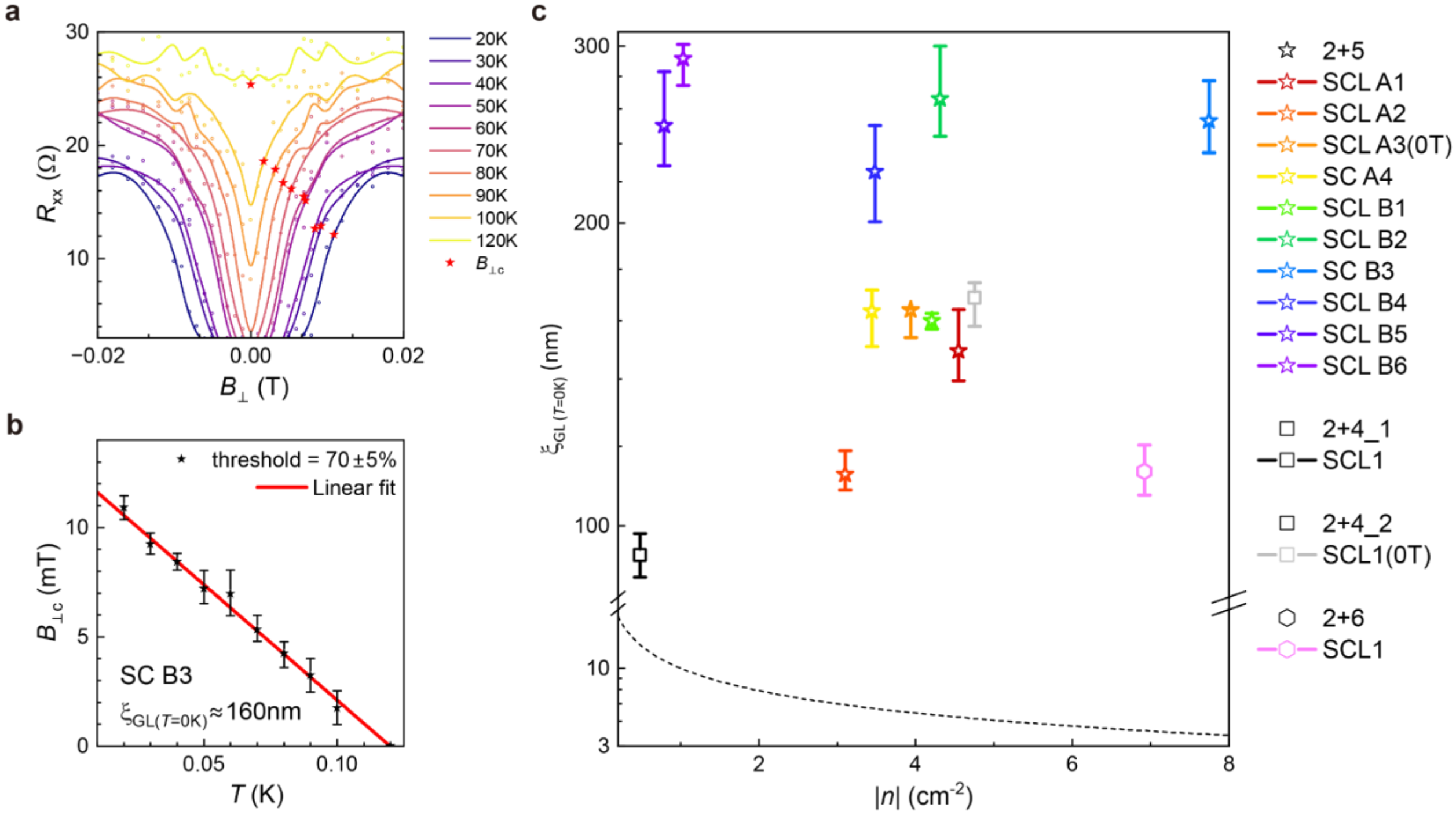

**Fig. 5 | Large Ginzburg-Landau coherence length of SC states. a,** Longitudinal resistance $R_{xx}$ of SC B3 in Device 2+5 as a function of perpendicular magnetic field $B_{\perp}$ at selected temperatures, measured at $B_{\parallel}$ = 1 T. Open circles denote the measured data and coloured lines are LOESS-smoothed traces. Filled red stars mark the critical perpendicular fields $B_{\perp c}$, defined by $R_{xx}$=0.7$R_N$, where $R_N$ is the normal-state resistance. **b,** Temperature dependence of $B_{\perp c}$ extracted from **a.** Linear fits to the critical magnetic field at SC B3 for $B_{\parallel}$ = 1 T. Error bars indicate the range obtained using resistance thresholds of 65% and 75% of $R_N$. The red line represents the linear fit to the Ginzburg–Landau form, yielding a zero-temperature coherence length $\xi_{GL(T=0K)}$ ≈ 160 nm. **c,** Zero-temperature Ginzburg–Landau coherence lengths $\xi_{GL(T=0K)}$ extracted from linear fits for representative SC states across the (2+n) devices. Data points correspond to the 70% resistance criterion, with error bars determined from the 65% and 75% criteria. The black dashed line denotes the mean inter-carrier distance $d_{particle}$=$|n|^{-1/2}$. The vertical axis is logarithmic. Unless otherwise indicated, measurements were performed at $B_{\parallel}$ = 1 T; SC A3 in Device 2+5 and SC1 in Device 2+4_2 were measured at $B_{\parallel}$ = 0 T.

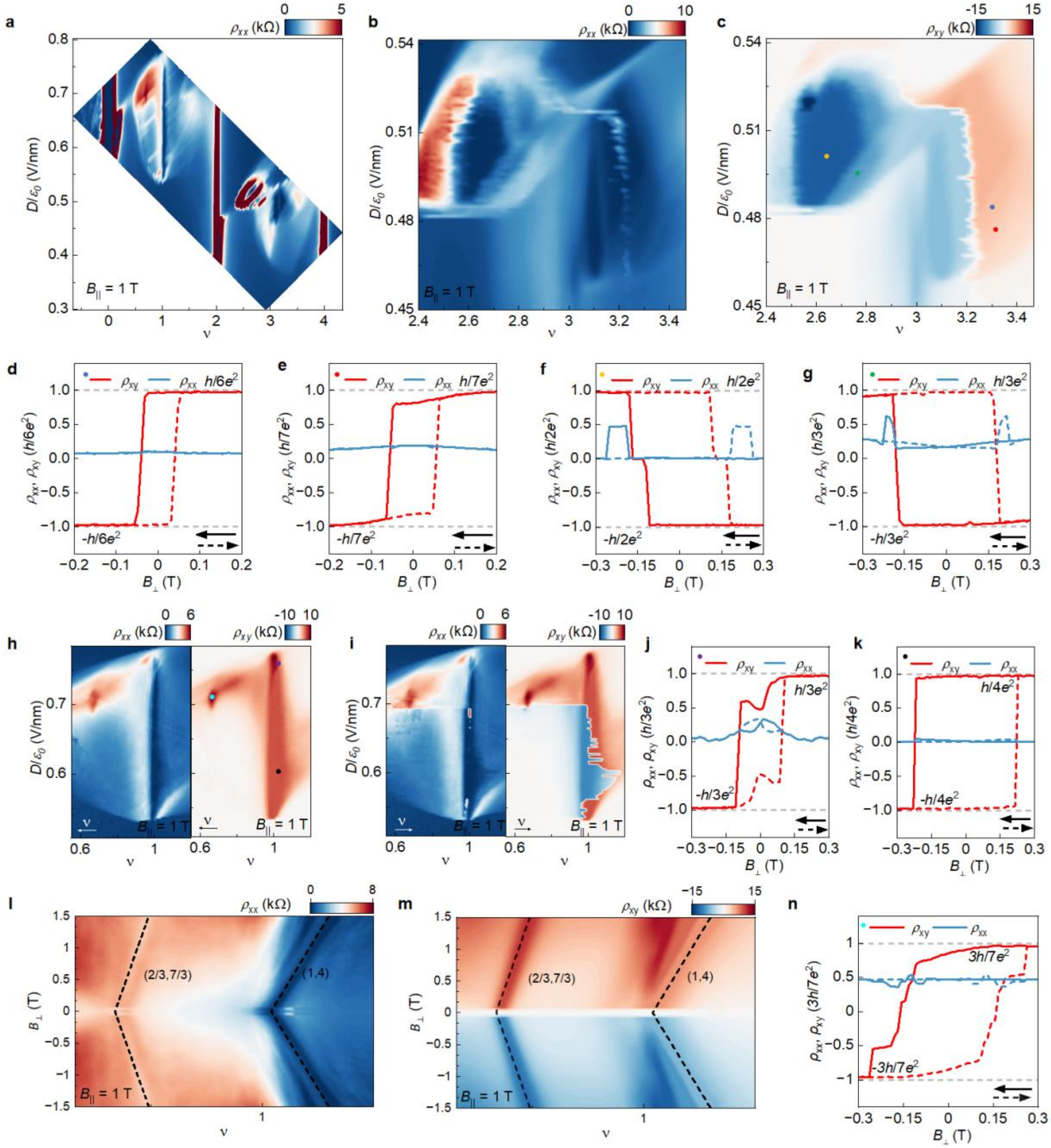

**Extended Data Fig. 1 | Robust HCIs in Device 2+4_1 under $B_{\parallel} = 1$ T. a,** Longitudinal resistivity $\rho_{xx}$ of Device 2+4_1 ($\theta = 1.38°$) as a function of filling factor $\nu$ and displacement field $D$ at $B_{\parallel} = 1$ T. **b,c,** Longitudinal $\rho_{xx}$ (**b**) and Hall $\rho_{xy}$ (**c**) resistivity as functions of $\nu$ and $D$ near $\nu = 3$ at $B_{\parallel} = 1$ T, resolving HCI states with different Chern numbers. **d–g,** Magnetic hysteresis loops of $\rho_{xx}$ and $\rho_{xy}$ as a function of perpendicular magnetic field $B_{\perp}$ measured at the representative locations indicated in **b**. **h,i,** $\rho_{xx}$ (**h**) and $\rho_{xy}$ (**i**) as functions of $\nu$ and $D$ near $\nu = 1$ measured for opposite carrier-density sweep directions at $B_{\parallel} = 1$ T. The C = 4 state exhibits opposite Hall polarities for the two sweep directions. **j,k,** Magnetic hysteresis loops of $\rho_{xx}$ and $\rho_{xy}$ for the C = 3 (**j**) and C = 4 (**k**) HCI states near $\nu = 1$. **l,m,** Landau fan diagrams of symmetrized $\rho_{xx}$ (**l**) and antisymmetrized $\rho_{xy}$ (**m**) along $D/\varepsilon_0 = 0.702$ V/nm at $B_{\parallel} = 1$ T. Black dashed lines mark the C = 4 trajectory from $\nu$

= 1 and the C = 7/3 trajectory from $\nu$ = 2/3.**n,** Magnetic hysteresis loops of $\rho_{xx}$ and $\rho_{xy}$ for the C = 7/3 FCI states near ν = 2/3. These robust behaviors demonstrate the persistence of Chern insulators at $B_{||}$ = 1 T.

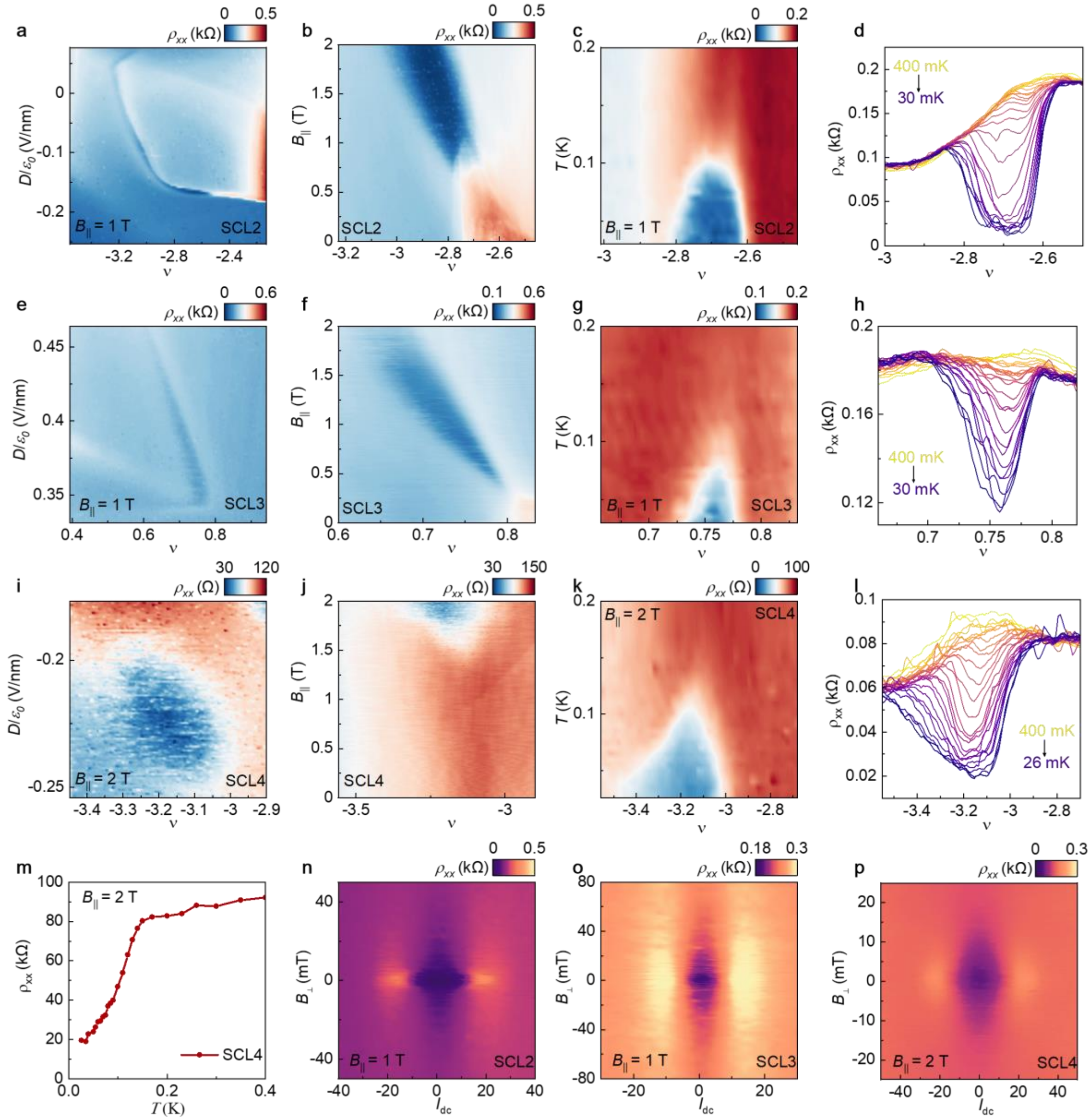

**Extended Data Fig. 2 | Characterization of additional SC states in Device 2+4_1. a-d,** Characterization of SCL2 at $B_{||}$ = 1 T. **a,** Longitudinal resistivity $\rho_{xx}$ as a function of $\nu$ and $D$. **b,** $\rho_{xx}$ as a function of $\nu$ and $B_{||}$. **c,** Temperature-evolution of SCL2, measured as $\rho_{xx}$ versus $\nu$ and $T$. **d,** Representative resistivity traces as functions of $\nu$ at selected $T$. **e–h,** Corresponding measurements of SCL3, arranged as in **a–d** at $B_{||}$ = 1 T. **i–l,** Corresponding measurements of SCL4 at $B_{||}$ = 2 T. **m,** Temperature dependence of $\rho_{xx}$ for SCL4. **n–p,** Differential resistance d$V$/d$I$ as a function of DC bias current $I_{DC}$ and perpendicular magnetic field $B_{\perp}$ for SCL2 (**n**), SCL3 (**o**) and SCL4 (**p**).

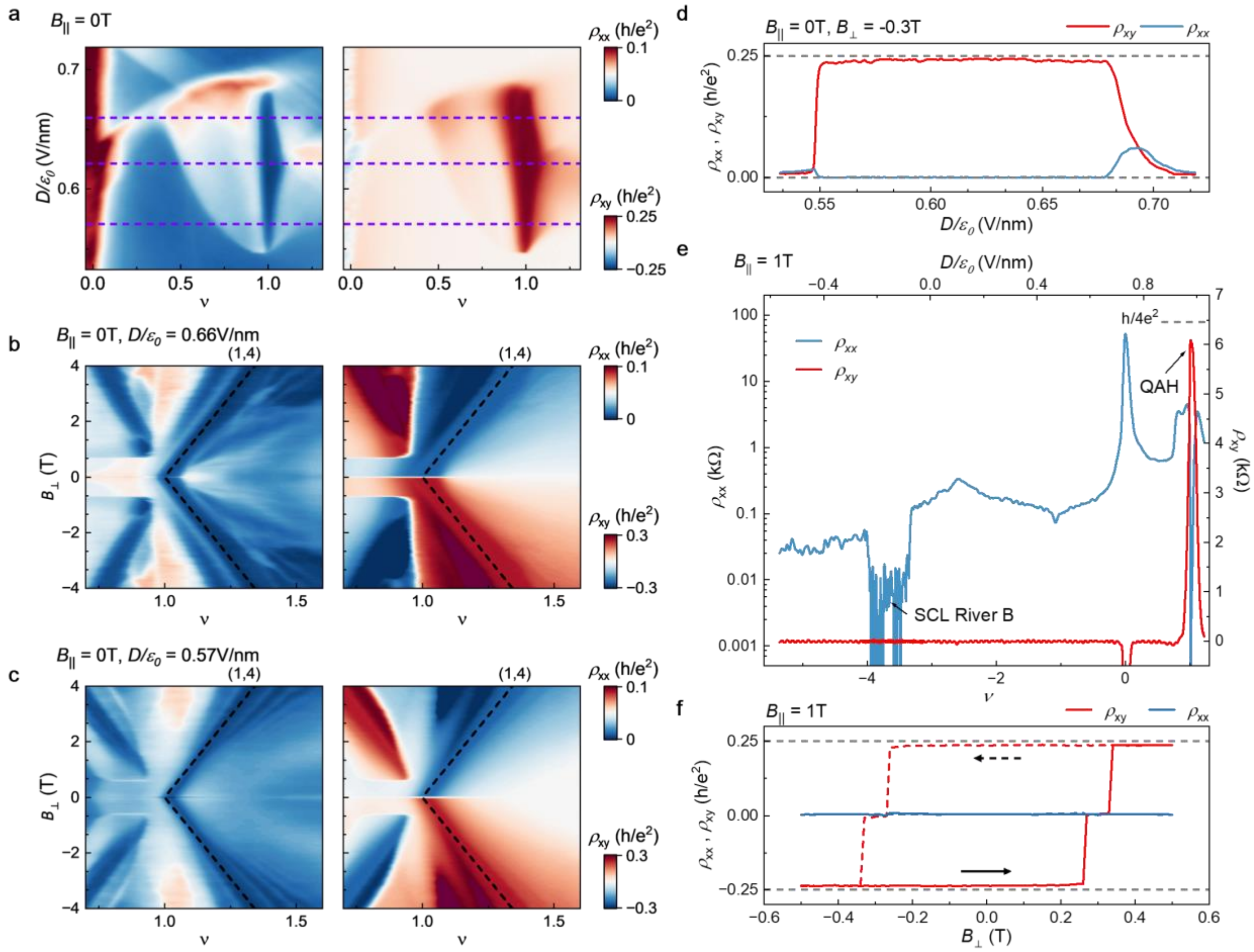

**Extended Data Fig. 3 | Characterization of the C = 4 HCI in Device 2+5. a,** Symmetrized longitudinal resistivity $\rho_{xx}$ (left) and antisymmetrized Hall resistivity $\rho_{xy}$ (right) as functions of $v$ and $D$, obtained from measurements at $B_\perp = \pm 0.3$ T. **b,c,** Landau fan diagrams of $\rho_{xx}$ and $\rho_{xy}$ at $D/\varepsilon_0$ = 0.66 V/nm and $D/\varepsilon_0$ = 0.57 V/nm. Black dashed lines indicate the C = 4 trajectory emanating from $v = 1$ according to the Středa formula. **d,** $\rho_{xx}$ and $\rho_{xy}$ as functions of $D$ at $v = 1$ and $B_\perp$ = -0.3 T. **e,** $\rho_{xx}$ and $\rho_{xy}$ along a representative trajectory at $B_{||}$ = 1 T, crossing SC River B on the hole-doped side and the C = 4 QAH state near $v = 1$. **f,** Magnetic hysteresis loop of $\rho_{xx}$ and $\rho_{xy}$ at $v$ = 1and $B_{||}$ = 1 T, with $\rho_{xy}$ reaches 95% of h/4e².

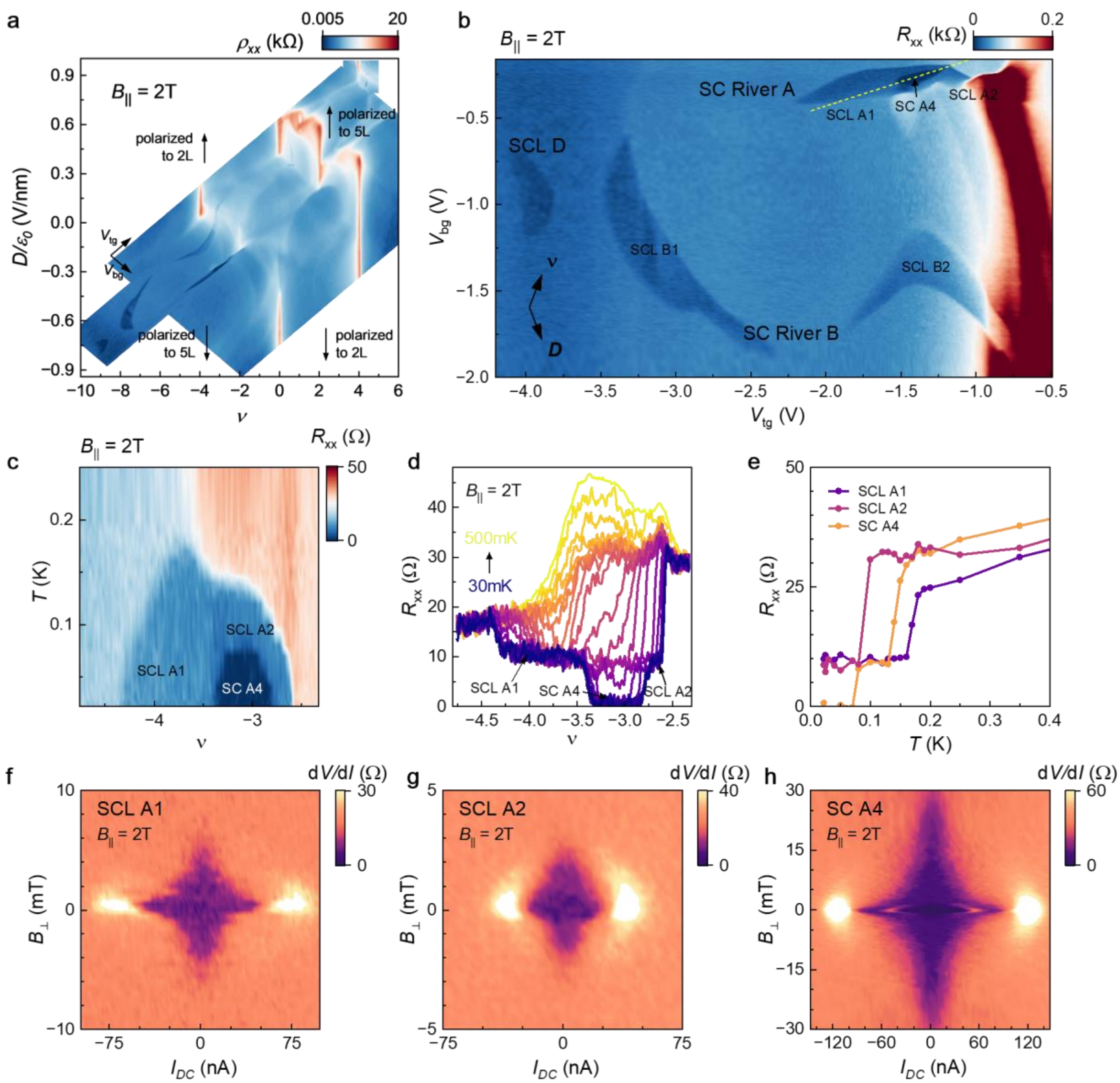

**Extended Data Fig. 4 | SC characterization in Device 2+5 at $B_{||}$ = 2 T. a,** Longitudinal resistivity $\rho_{xx}$ of Device 2+5 as a function of $\nu$ and $D$ at $B_{||}$ = 2T. **b,** Longitudinal resistance $R_{xx}$ on the hole-doped side as a function of dual-gate voltages at $B_{||}$ = 2 T. **c,** Temperature evolution of SCL River A along the yellow dashed trajectory in **b**, measured as $R_{xx}$ versus $\nu$ and T. **d,** $R_{xx}$ as a function of $\nu$ at selected $T$. **e,** Temperature dependence of $R_{xx}$ for SC A. **f-h,** Differential resistance d$V$/d$I$ as functions of $I_{DC}$ and $B_{\perp}$ for SCL A1 (**f**), A2 (**g**), and SC A4 (**h**) at $B_{||}$ = 2 T.

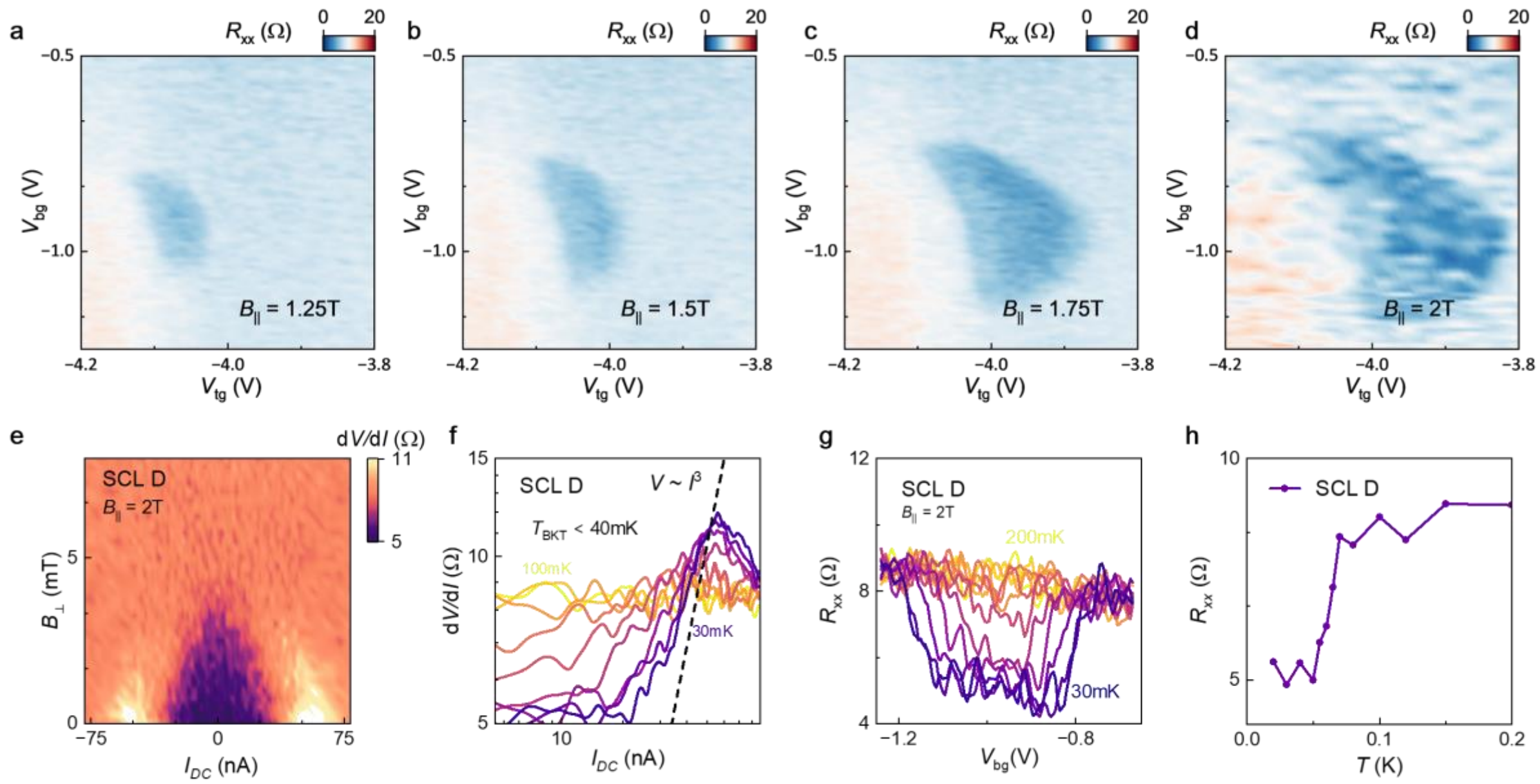

**Extended Data Fig. 5 | Characterization of the high-$B_{\parallel}$-induced SCL D in Device 2+5. a-d,** Longitudinal resistance $R_{xx}$ as a function of dual-gate voltages in the vicinity of SCL D at $B_{\parallel}$ = 1.25T (**a**), 1.5T (**b**), 1.75T (**c**), 2T (**d**). The spatially irregular background in **d** arises from measurement noise. **e,** Differential resistance d$V$/d$I$ as a function of $I_{DC}$ and $B_{\perp}$ for SCL D at $B_{\parallel}$ = 2 T. **f,** Temperature dependent d$V$/d$I$ traces as functions of $I_{DC}$ at $B_{\parallel}$ = 2 T. The $V_{xx} \sim I_{DC}^3$ criterion places the BKT transition below approximately 40 mK, comparable to the estimated minimum electron temperature. **g,** $R_{xx}$ as a function of bottom-gate voltage $V_{bg}$ at selected $T$ at $B_{\parallel}$ = 2 T. **h,** Temperature dependence of $R_{xx}$ or SCL D, yielding $T_c \approx 56$ mK using a 70% normal-state-resistance criterion.

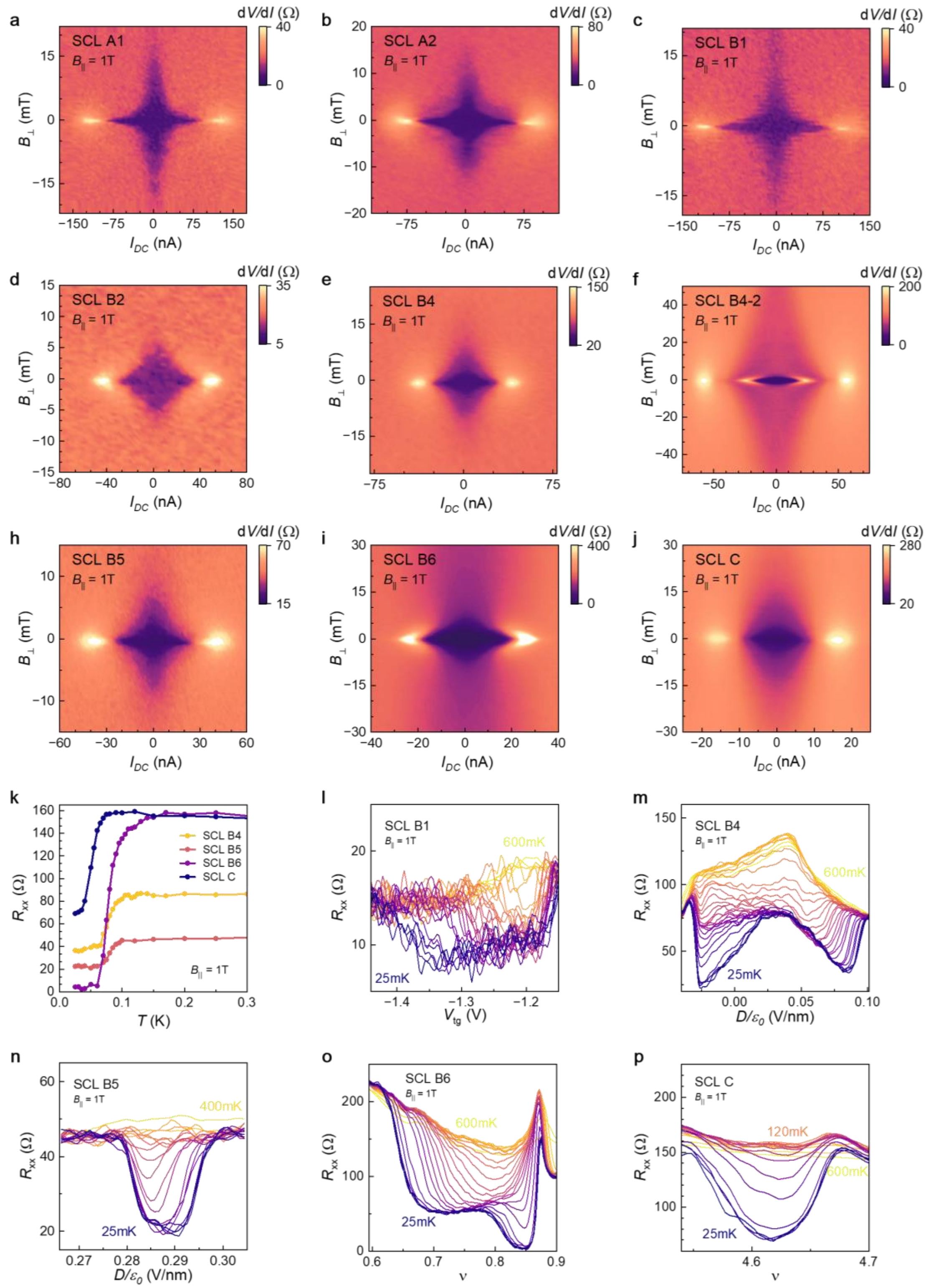

**Extended Data Fig. 6 | Additional characterization of SC states in Device 2+5 at $B_\parallel = 1$ T. a-j,** Differential resistance d$V$/d$I$ as a function of $I_{DC}$ and $B_\perp$ at various SC positions, acquired under

an in-plane field of $B_{||}$ = 1 T. SCL B4-2 denotes a different position within the SC B4 region ($V_{tg}$ = -0.77 V, $V_{bg}$ = -1.62V). **k,** Temperature dependence of $R_{xx}$ for SCL B4–B6 and C. **l–p,** Temperature evolution of $R_{xx}$ for SCL B1 (**l**), B4 (**m**), B5 (**n**), B6 (**o**) and C (**p**).

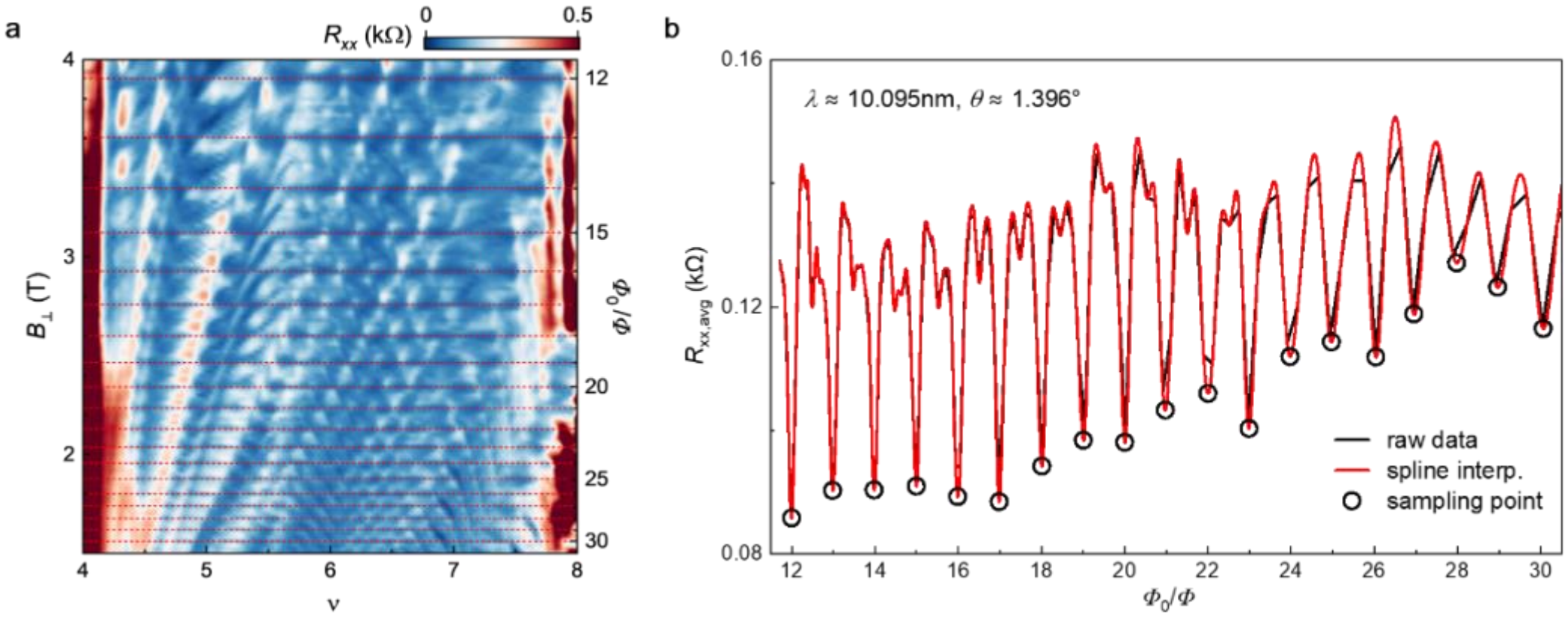

**Extended Data Fig. 7 | Representative Brown–Zak oscillations in Device 2+5. a,** Longitudinal resistance $R_{xx}$ as a function of $\nu$ and $B_\perp$ at $D/\varepsilon_0$ = 0 V/nm. Red dashed lines mark the Brown–Zak oscillations, which become approximately equally spaced when expressed in terms of $N = \Phi_0/\Phi$ ($\Phi = A_m \cdot B_\perp$, the magnetic flux through one moiré unit cell). **b,** Filling averaged $R_{xx,avg}$ as a function of $N$, averaged over $4.5 < \nu < 6.5$. The black trace shows the raw data, while the red trace is a spline interpolation. The circles indicate the oscillation positions identified in **a**. A linear fit to these circles gives oscillation period $\Delta(1/B_\perp)$, enabling extraction of the moiré wavelength $\lambda$ and twist angle $\theta$.

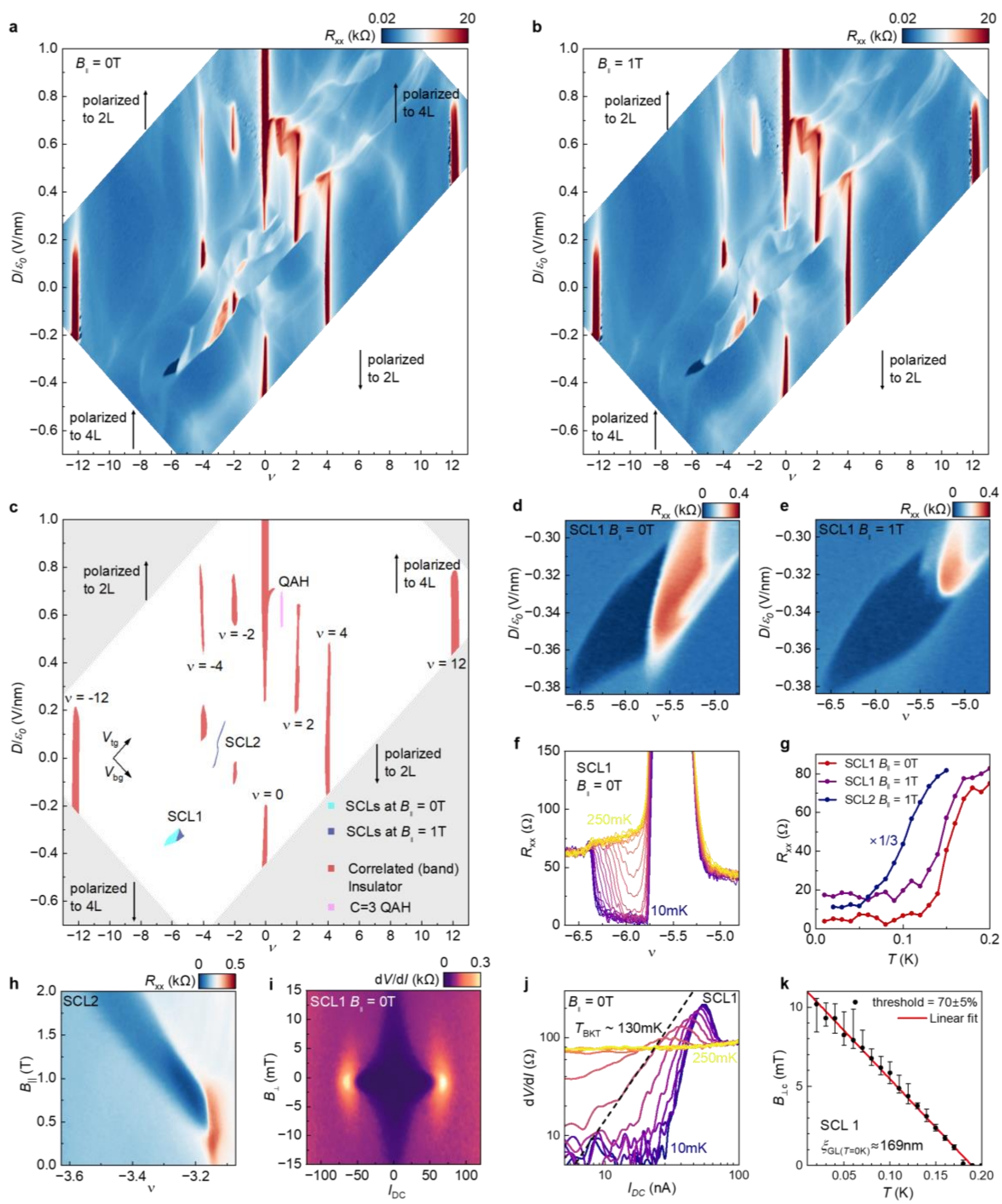

**Extended Data Fig. 8 | SC and HCI states in Device 2+4_2 with $\theta$ = 1.18°. a,b,** Longitudinal resistance $R_{xx}$ as a function of $\nu$ and $D/\varepsilon_0$ measured at $B_\parallel$ = 0T (**a**) and $B_\parallel$ = 1T (**b**) at $T$ = 10mK. **c,** Schematic $\nu$–$D$ phase diagram summarizing the principal superconducting, QAH and correlated insulating states. of Device 2+4_2 ($\theta$ = 1.18°). Gray reigions mark the experimentally inaccessible regime. The arrows indicate the directions of the dual-gate tuning directions. **d,e,** Enlarged $R_{xx}$ maps as around SCL1 at $B_\parallel$ = 0T (**d**) and $B_\parallel$ = 1T (**e**). **f,** $R_{xx}$ as a function of $\nu$ at $D/\varepsilon_0$ = -0.345 V/nm for temperatures from 10 to 250 mK from for SCL1 at $B_\parallel$ = 0T. **g,** Temperature dependence

of $R_{xx}$ at $\nu$ = -6.03, $D/\varepsilon_0$ = -0.345 V/nm for SCL1 at $B_\parallel$ = 0T and $B_\parallel$ = 1T, $\nu$ = -3.30, $D/\varepsilon_0$ = -0.022 V/nm for SCL2. **h,** Evolution of SCL2 with in-plane magnetic field, measured as $\rho_{xx}$ versus $\nu$ and $B_\parallel$ at $D/\varepsilon_0$ = -0.022 V/nm. **i,** Differential resistance d$V$/d$I$ as a function of bias current $I_{DC}$ and $B_\perp$ at $\nu$ = -5.86, $D/\varepsilon_0$ = -0.339 V/nm for SCL1 at $B_\parallel$ = 0T, exhibiting a critical current of about 50 nA. **j,** Logarithmic plot of d$V$/d$I$ as a function of $I_{DC}$ at $\nu$ = -6.03, $D/\varepsilon_0$ = -0.345 V/nm for SCL1 at $B_\parallel$ = 0T. The black dashed line marks the $V_{xx} \propto I_{DC}^3$ criterion, which yields $T_{BKT}$ ~ 130mK. **k,** Temperature dependence of critical perpendicular magnetic field at various temperatures at $\nu$ = -5.86 , $D/\varepsilon_0$ = -3.39 V/nm for SCL1. The straight line is a fit to the Ginzburg–Landau equation, which yields coherent length $\xi_{GL} \approx$ 169 nm.

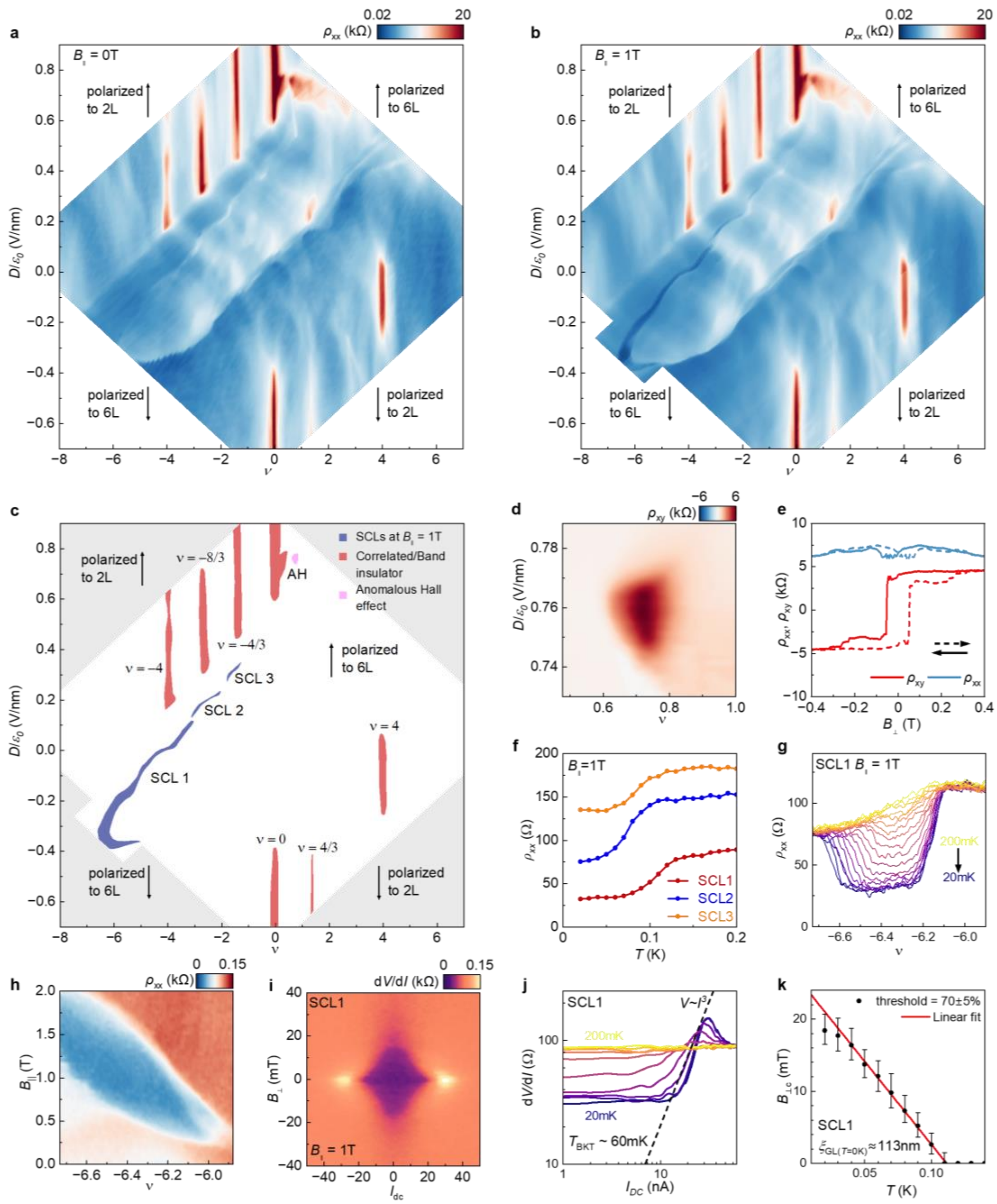

**Extended Data Fig. 9 | SC and anomalous Hall effect in Device 2+6 ($\theta$ = 1.37°). a,b,** Longitudinal resistivity $\rho_{xx}$ as a function of $\nu$ and $D/\varepsilon_0$ measured at in-plane magnetic field $B_\parallel$ = 0T **(a)** and $B_\parallel$ = 1T **(b)** with temperature $T$ = 20mK of twisted 2+6 graphene device. A SC region develops with applied in-plane magnetic field. **c,** Schematic $\nu$–$D$ phase diagram summarizing the principal superconducting, anomalous Hall and correlated insulating states. Gray regions marks the experimentally inaccessible regime. The directions of the dual gates are labeled by the arrows. **d,** Anti-symmetrized Hall resistivity $\rho_{xy}$ **(b)** as a function of $\nu$ and $D/\varepsilon_0$ measured at perpendicular magnetic field $B_\perp$ = ±0.2 T around $\nu$ = 2/3. **e,** Magnetic hysteresis loop of symmetrized $\rho_{xx}$ and anti-symmetrized $\rho_{xy}$ as functions of $B_\perp$ at $\nu$ = 2/3, $D/\varepsilon_0$ = 0.758 V/nm. The arrows indicate sweeping directions. **f,** $\rho_{xx}$ as a function of temperature at $\nu$ = -6.36, $D/\varepsilon_0$ = -0.315 V/nm for SCL1, $\nu$ = -2.78, $D/\varepsilon_0$ = 0.168 V/nm for SCL2, and $\nu$ = -1.48, $D/\varepsilon_0$ = 0.312 V/nm for SCL3. **g,** $\rho_{xx}$ as a function of $\nu$ at $D/\varepsilon_0$ = -0.335 V/nm for temperature ranging from 20mK to 200mK. **h,** $\rho_{xx}$ as a function of $\nu$ and $B_\parallel$ at $D/\varepsilon_0$ = -0.335 V/nm. **i,** Differential resistance d$V$/d$I$ as a function of bias current $I_{DC}$ and $B_\perp$ at $\nu$ = -6.36, $D/\varepsilon_0$ = -0.315 V/nm for SCL1, exhibiting a critical current of about 20 nA. **j,** Logarithmic plot of d$V$/d$I$ as a function of $I_{DC}$. The black dashed line marks the $V \propto I_{DC}^3$ criterion, which yields BKT transition temperature $T_{BKT}$ ~ 60mK. **k,** Temperature dependence of critical perpendicular magnetic field for SCL1. The red line is a fit to the Ginzburg–Landau equation, which yields coherent length $\xi_{GL}$ ≈ 113nm.

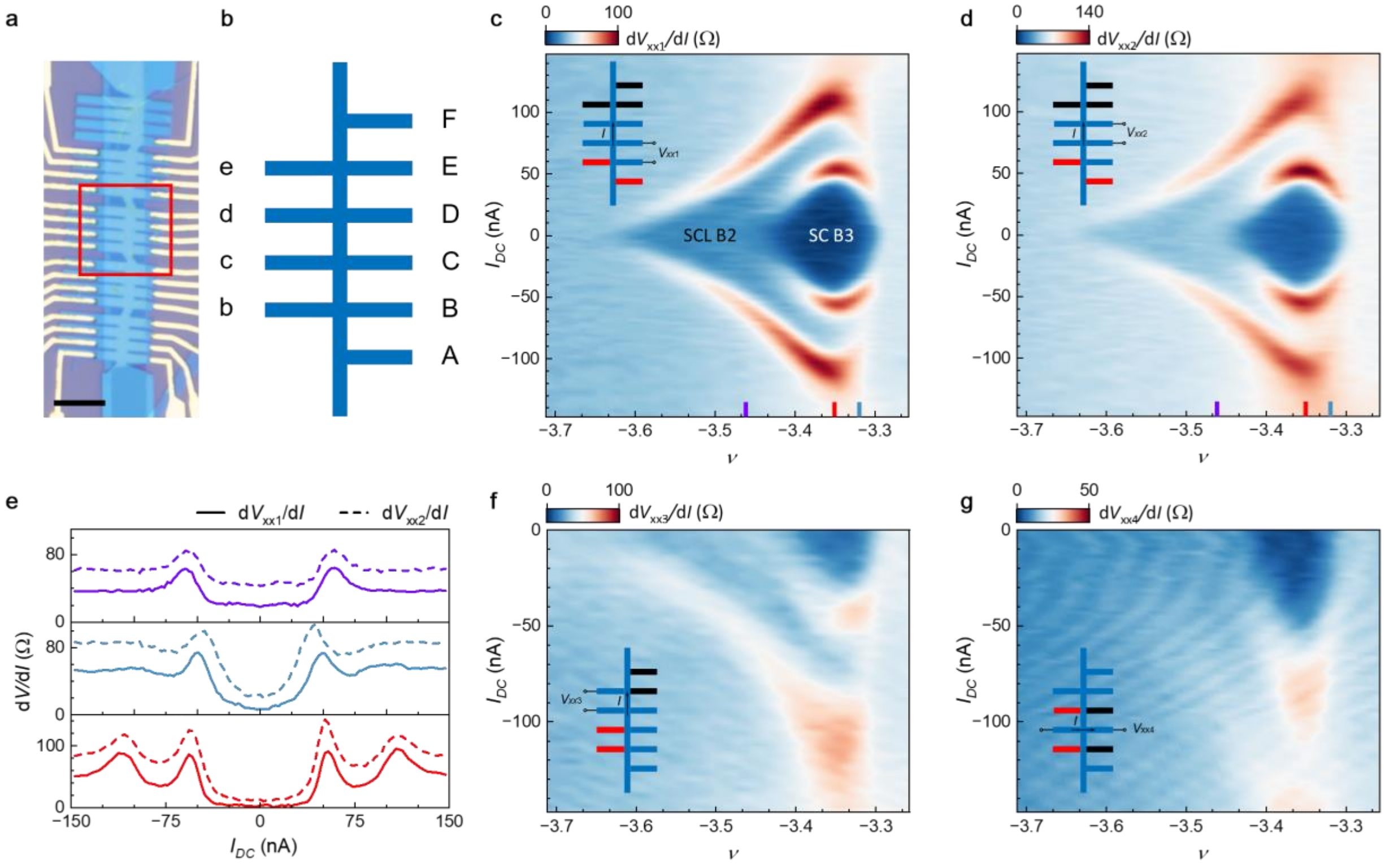

**Extended Data Fig. 10 | Transport characterization of spatial inhomogeneity in Device 2. a,** Optical micrograph of Device 2+5. The red rectangle marks the region for transport measurements, with a black scale bar of 10 μm. **b,** Schematic of the measurement geometry. **c,d,** Differential resistance d$V$/d$I$ as a function of filling factor $\nu$ and bias current $I_{DC}$, measured with current ($I_{AC}$ = 3nA) applied between contacts A and b, and contacts e&E&F grounded. The four-terminal voltage was measured between contacts B and C (**c**), and between C and D (**d**). **e,** Representative nonlinear

differential-resistance traces extracted from **c,d** at selected filling factors. The purple, blue and red curves correspond to $\nu$ = -3.46, -3.32 and -3.35, respectively. **f,** d$V$/d$I$ as a function of $\nu$ and $I_{DC}$, measured with current ($I_{AC}$ = 3nA) applied between contacts b and c, and contacts E&F grounded. The four-terminal voltage was measured between contacts d and e. **g,** d$V$/d$I$ as a function of $\nu$ and $I_{DC}$, measured with current ($I_{AC}$ = 20nA) applied between contacts b and d, and contacts B and D grounded. The four-terminal voltage was measured between contacts c and C, yielding an approximately transverse transport configuration. All measurements were performed at $D/\varepsilon_0$ = -0.11 V/nm. Similar nonlinear d$V$/d$I$ features are observed for all measurement configurations.

| Device | Twist angle ($\theta$) | HCI or FHCI states | $B_{\parallel}$ response (0~2T) of SC states | Pauli-limit violation | Zero-$B_{\parallel}$ field SC |
|---|---|---|---|---|---|
| D 2+4_1 | 1.38° | C = 3 and 4 at $\nu$ = 1<br>\|C\| = 1-7 around $\nu$ = 3<br>C = 7/3 at $\nu$ = 2/3 | SCL 1, induced → enhanced<br>SCL 2, induced → enhanced/suppressed<br>SCL 3, induced → enhanced → suppressed<br>SC 4, induced | Yes | No |
| D 2+4_2 | 1.18° | C = 3 at $\nu$ = 1 | SCL 1, enhanced → suppressed<br>SCL 2, induced → enhanced | Yes | Yes |
| D 2+5 | 1.40° | C = 4 at $\nu$ = 1 | SCL River A and B, enhanced/suppressed/induced<br>SCL C and D, induced → enhanced/suppressed | Yes<br>(except SCL A3 and B7) | Yes |
| D 2+6 | 1.37° | AHE | SCL 1, induced→ enhanced<br>SCL 2, induced → enhanced → suppressed<br>SCL 3, induced → suppressed | Yes | No |

**Extended Data Table 1 | Summary of SC and topological states across the twisted (2+n) RMG devices.** The table summarizes the twist angle, observed HCI or fractional HCI states, and the evolution of the SC states under $B_{\parallel}$ from 0 to 2 T. “Induced”, “enhanced” and “suppressed” denote the emergence, strengthening and weakening of SC with increasing $B_{\parallel}$, respectively. Arrows indicate the sequential evolution of each SC state with increasing $B_{\parallel}$. Pauli-limit violation indicates whether the observed SC states persist beyond the estimated weak-coupling Pauli limit; exceptions are specified individually. The final column indicates whether SC is already present at $B_{\parallel}$=0. In all measurements, zero-resistance behaviour is observed in SC A4, SC B3 and portions of SCL A2, A3 and B6 in Device 2+5.